\documentclass{aa}

\usepackage{graphicx,amssymb,amsmath}       
\usepackage{natbib}
\usepackage{longtable}
\usepackage{xtab,booktabs}
\usepackage{caption}
\usepackage{numprint}

\def\H0{{\rm ~km~s^{-1}~Mpc^{-1}}}

\def\.25{0.25 keV\thinspace}

\def\d19{D$\,\leq\,$19~Mpc} 
\def\dgtr19{D$\,>\,$19~Mpc}

\let\lesssim=\la
\let\gtrsim=\ga

\begin{document}

\title{15-GHz radio emission from nearby low-luminosity active galactic nuclei}

\author{
   Payaswini Saikia
   \inst{1,2}
   \and
   Elmar K\"{o}rding
   \inst{1}
   \and 
   Deanne L. Coppejans
   \inst{3}
    \and
      Heino Falcke
   \inst{1,4}
   \and
    David Williams
    \inst{5}
    \and\\
    Ranieri D. Baldi
    \inst{5}
    \and
   Ian Mchardy
   \inst{5}
    \and
    Rob Beswick
    \inst{6}
}
\institute{Department of Astrophysics/IMAPP, Radboud University, Nijmegen, 6500 GL Nijmegen, The Netherlands \\
           \email{p.saikia@astro.ru.nl} 
           \and
           New York University Abu Dhabi, PO Box 129188, Abu Dhabi, UAE
           \and
           Center for Interdisciplinary Exploration and Research in Astrophysics (CIERA) and Department of Physics and Astronomy, Northwestern University, Evanston, IL 60208
           \and
           Netherlands Institute for Radio Astronomy (ASTRON), PO Box 2, 7990 AA Dwingeloo, The Netherlands
           \and
           Department of Physics and Astronomy, University of Southampton, Highfield, Southampton SO17 1BJ
           \and
           Jodrell Bank Centre for Astrophysics, School of Physics and Astronomy, The University of Manchester, Manchester, M13 9PL, UK
                     }

\date{May 2018}

\abstract{
We present a sub-arcsec resolution radio imaging survey of a sample of 76 low-luminosity active galactic nuclei (LLAGN) that were previously not detected with the Very Large Array at 15 GHz. Compact, parsec-scale radio emission has been detected above a flux density of 40 $\mu$Jy in 60 \% (45 of 76) of the LLAGN sample. We detect 20 out of 31 (64 \%) low-ionization nuclear emission-line region (LINER) nuclei, ten out of 14 (71 \%) low-luminosity Seyfert galaxies, and 15 out of 31 (48 \%) transition objects. We use this sample to explore correlations between different emission lines and the radio luminosity. We also populate the X-ray and the optical fundamental plane of black hole activity and further refine its parameters. We obtain a fundamental plane relation of log L$_\textrm{R}$ = $0.48\,(\pm0.04$) log L$_\textrm{X}$ + $0.79\,(\pm0.03$) log $\textrm{M}$ and an optical fundamental plane relation of log L$_\textrm{R}$ = $0.63\,(\pm 0.05)$ log L$_{[\rm O~III]}$  + $0.67\,(\pm 0.03)$ log $\textrm{M}$ after including all the LLAGN detected at high resolution at 15 GHz, and the best-studied hard-state X-ray binaries (luminosities are given in erg s$^{-1}$ while the masses are in units of solar mass). Finally, we find conclusive evidence that the nuclear 15 GHz radio luminosity function (RLF) of all the detected Palomar Sample LLAGN has a turnover at the low-luminosity end, and is best-fitted with a broken power law. The break in the power law occurs at a critical mass accretion rate of 1.2$\times$10$^{-3}$ M$_{\odot}$/yr, which translates to an Eddington ratio of $\rm \dot m_{Edd} \sim 5.1 \times 10^{-5}$, assuming a black hole mass of 10$^9 M_{\odot}$. The local group stands closer to the extrapolation of the higher-luminosity sources, and the classical Seyferts agree with the nuclear RLF of the LLAGN in the local universe.

} % end of abstract

\titlerunning{Radio Nuclei in Palomar AGN}
\authorrunning{Saikia P. et al. 2018}

\maketitle

\section{Introduction}
\label{secintro}

Accretion and jet formation in active galactic nuclei (AGN) have been extensively studied in bright systems like quasars and Fanaroff-Riley Type 1 and Type 2 \citep[FR1 and FR2;][]{fr} radio galaxies. Low-luminosity AGN (LLAGN, AGN with L$_{\rm bol} <$10$^{42}$ erg s$^{-1}$) are fainter AGN accreting at very low Eddington rates. These sources are significantly more numerous than the higher-luminosity AGN. According to the Palomar spectroscopic survey of nearby galaxies \citep{h97}, more than 40\% of the galaxies in the local universe are comprised of LLAGN. However, they are intrinsically faint, and thus are neither as well-observed nor as well-understood as the brighter AGN classes like quasars and FR galaxies.

Observational and theoretical studies of LLAGN suggest that although these objects differ from their high-power cousins, they are also powered by accretion onto a black hole \citep[eg.,][]{heino2000}. Most of these sources have low-ionization state spectra \citep[eg.,][]{h97}. They seem to have radiatively inefficient accretion flows \citep[eg.,][]{ho2008}. The typical spectral energy distribution of a LLAGN has an Infra-red excess, while the big blue bump in the optical and ultra-violet (UV) region (typical in a quasar spectra) is generally absent \citep[eg.,][]{hoet2010}. They are mostly seen to be radio-loud compared to the general AGN population \citep[eg.,][etc.]{n05,sikora2007} and have radio emission indicating the presence of compact radio jets \citep[eg.,][etc.]{fb99,f2001,n05}. The X-ray spectra are generally described by a simple power law with $\Gamma \sim$ 1.7 - 1.9, and the 6.4 keV Fe K$\alpha$ is almost always narrow \citep[eg.,][]{ter2002}.

Physical as well as statistical properties of the LLAGN are extremely important to understand as they bridge the physical parameter-space between the supermassive black holes in active and normal galaxies like our own Sgr A$^{\star}$. The most studied LLAGN sample to date is the optically-selected Palomar spectroscopic survey of all ($\sim$488) northern galaxies \citep{h95}. Radio detections of this sample however, are severely limited in the high frequency range ($\ge$ 15 GHz). These are required to observe and study the compact, nuclear core of the relativistic jet. The most complete study of the Palomar LLAGN sample at 15 GHz to date has a detection rate of 35\%, with a formal detection limit of 1.5 mJy \citep{n05}.

A complete radio sample of LLAGN is necessary to answer some of the key questions regarding the physics of AGN and their host galaxies, such as what is the shape of the radio luminosity function of AGN at the lowest luminosities, how is the accretion disk coupled to its jet, what is the effect of the host galaxy or the black hole spin on the jet and accretion power, does accretion physics scale with black hole mass, and do LLAGN belong to the analogue of the hard-state X-Ray Binaries (XRBs).

In this paper, we present a high resolution ($\sim$0.2 arcsec) and high frequency (15 GHz) radio imaging survey of 76 previously undetected LLAGN with the National Science Foundation's (NSF's) Karl G. Jansky Very Large Array \citep[VLA,][]{vla11} in A-configuration. Radio emission from LLAGN is thought to originate in a compact jet, which produces a flat spectrum radio core with high brightness temperature \citep[eg.,][]{heino2000}. Observations with the high spatial resolution of the VLA at 15 GHz in A-configuration ensure that we can distinguish between the nuclear and non-nuclear emission, as any emission from star formation will be suppressed at such a high frequency. Moreover, observations at 15 GHz are unaffected by the obscuration present at UV or optical wavelengths and are less affected by free-free absorption compared to observations at longer centimeter wavelengths.

In Sect.~\ref{secsample}, we define the sample used and summarize the previous archival 15 GHz observations of the LLAGN sources in the Palomar sample. The new observations and the data reduction process are described in Sect. 3. The results of the observations are presented in Sect.~\ref{secres}. In Sect.~\ref{thefp}, we use the new radio detections to shed light on the mass-scaling relations of the central black holes. We refine the X-ray and optical fundamental plane of black hole activity, which is a relationship between the radio luminosity, OIII/X-ray luminosity and the black hole mass of both hard-state stellar-mass black holes and their supermassive analogs. While X-ray emission is a more direct proxy for the accretion rate of an active galaxy, [OIII] line luminosity is relatively easier to obtain, it can be measured by ground-based observations, and the contamination coming from relativistic beaming and the effects of torus obscuration are minimized. However, as it is an indirect tracer of accretion rate, [OIII] emission introduces more scatter to the fundamental plane. We discuss both these planes in Sect.~\ref{thefp}. The radio luminosity function of the LLAGN sample is discussed in Sect.~\ref{therlf}. We finally discuss various physical properties of the LLAGN sample and highlight the main conclusions of this study in Sect.~\ref{secconclusion}. In this paper, we use a Hubble constant H$_0\,= 75$ km s$^{-1}$ Mpc$^{-1}$ to be consistent with \citet{h97} who tabulate the results of optical spectroscopy of the Palomar sample.

\section{Sample}
\label{secsample}

The sample of LLAGN studied in the paper is taken from the optically-selected Palomar spectroscopic survey of all ($\sim$488) northern galaxies with B$_\tau$ $<$ 12.5 mag \citep{h95}. The spectroscopic parameters as well as the nuclear activity classification of 418 galaxies in the Palomar spectroscopic survey, which show nuclear emission lines, were presented in \citet{h97}. Of these 418 galaxies, we consider only the 403 that belong to the defined Palomar sample. It is comprised mainly of low-luminosity Seyfert galaxies, low ionization nuclear emission region (LINER) objects, and transition nuclei. This sample represents the local galaxy population up to $\sim$ 120 Mpc. It has the largest database of homogeneous, highly sensitive and high-quality optical spectra of nearby galaxies, making it ideal to study the LLAGN in the nearby universe.

Of the 403 sources in the Palomar sample, 206 are HII region type spectra and the rest (197) are active galaxies (AGN and LLAGN). \cite{n05} observed all the 197 active sources in the Palomar sample with the VLA, and detected 68 of them. With the new improved, highly sensitive VLA, we observed 76 of the remaining 129 sources. These 76 sources were randomly selected, without any additional biases or constraints. We plan to observe the rest of the previously undetected sources (46 sources) in the next VLA A-configuration observing session and complete the sample. The 76 LLAGN reported in the paper consists of 31 pure LINERs, 31 transition objects, and 14 low-luminosity Seyfert galaxies. 

\subsection{Previous radio observations of the Palomar sample}

Several radio surveys have observed a substantial fraction of the the Palomar galaxies \citep[eg.,][etc.]{humet87, fabet89, caret90, heino2000, ulvho2001, houlv01, filho2004, n05, filho2006, panessa2013, lemm2018}. 

\cite{houlv01} did a 1$\arcsec$ resolution survey of the low-luminosity Seyferts in the Palomar sample, and detected over 80\% of the sources  with the VLA at 1.4 and 5 GHz. \cite{filho2002} did a similar survey (2.$\arcsec$5 resolution) for the composite LINER/HII galaxies in the Palomar sample with the VLA at 8.4 GHz, and detected radio cores in $\sim$25\% of the sources. Another survey of the Palomar LINERs and composite LINER/HII galaxies was performed by \cite{filho2006} with the high-resolution Multi-Element Radio-Linked Interferometer Network (MERLIN) at 5 GHz, and detected 21\% of the sources previously undetected in high resolution. \cite{panessa2013} performed a systematic radio survey of the radio-quiet Seyferts in the Palomar sample, at sub-pc scale, with very long baseline interferometry (VLBI) at 1.7 and 5 GHz. They found that out of 23 sources with a VLA detection, 17 are also detected with VLBI. \cite{lemm2018} did a high-resolution ($\le$0.2$\arcsec$) 1.5 GHz radio survey of 103 Palomar galaxies (both active and quiescent) with the eMERLIN array as part of the Legacy e-MERLIN Multi-band Imaging of Nearby Galaxies Survey (LeMMINGs), and detected radio emission for 47 of the 103 galaxies.

Prior to this work, the most complete radio survey of all the LLAGN and AGN in the Palomar sample at a uniform sensitivity and resolution was performed by \citet{n05}. They completed a sub-arcsecond resolution 15 GHz VLA survey of the Palomar galaxies, but they could detect only 35$\%$ of the complete sample due to the low sensitivity of the survey (rms$\sim$0.2 mJy/beam). The majority of the detected sources were found to have fluxes between 1.5 mJy and $\sim$ 300 mJy. 

\subsection{New radio observations of the Palomar sample}

In this study, we observed 76 of the previously undetected Palomar LLAGN at 15 GHz. With a total time of approximately four minutes per source using the 6.14 GHz bandwidth, we reached a rms noise level as low as 11.5 $\mu$Jy/beam, compared to previous surveys with the VLA at 15GHz that had noise levels of $\sim$0.3mJy/beam. The observations presented in this paper, together with the detection reported in \cite{n05}, comprise the most complete nuclear radio survey of the LLAGN and AGN in the Palomar sample at 15 GHz. 

\section{New VLA observations and data reduction}
\label{secobsvla}

The 76 previously undetected low-luminosity Seyferts and LINERs of the Palomar sample were observed at 15 GHz (2 cm) with the VLA over nine runs comprising a total of 11.42~hrs during October-December 2016. The VLA was in A-configuration at this time and was configured to observe in full polarization mode. There was 6.14 GHz bandwidth, divided into 48 spectral windows comprising 64 channels each.

Observations affected due to misapplication of the online atmospheric delay model affecting all VLA science observations from 9 August 2016 and 14 November 2016, were fixed by reprocessing the observations with the VLA calibration pipeline. As the on-source observing duration was short ($\sim$ 4-5 mins), the effect was minimal and could be fixed during re-processing. %The weather was fair with clear skies for most of the runs, except one run with a sky overcast (with mixed clouds) and 3 runs with a partial sky coverage (with stratiform or mixed clouds).

%The target sources were observed at elevations between 0{\degr} and 74{\degr}.

The target sources were observed with a $\sim$4-5 ~min integration time sandwiched between two $\sim$1~min observations on a nearby `phase calibrator'. Standard calibration was done with the VLA calibration pipeline with additional flagging. Subsequent imaging was done in CASA \citep{casa}, following the standard reduction procedures as outlined in the CASA cookbook, including cleaning of the sources, correcting for primary beam and fitting the sources to measure the flux densities of the sources. Observations of J1331+3030 (3C286) were used for most of the target sources to set the flux-density scale at 15~GHz. The flux calibrators J0137+3309 and J0542+4951 were used for the remaining 18 sources.

We reduced these data obtained from the different observing runs in the same way using the inbuilt pipeline in CASA (the Common Astronomy Software Applications package) version 4.6.0. We used PyBDSM \citep[Python Blob Detection and Source Measurement,][]{pybdsm} to measure the flux density of the sources and map them in Stokes I. The 1$\sigma$ error in flux, obtained by setting the flux density scale of the targets relative to the flux density calibrators, is expected to be roughly $\sim$2.5\%. The typical root mean square (rms) noise in the final maps is $\sim$ 11.5~$\mu$Jy/beam. % and we use a formal detection limit of 0.0~mJy (i.e. 5$\,\sigma$)

\section{Results of the VLA radio observations}
\label{secres}

\begin{figure}[t]
\resizebox{\columnwidth}{!}
{
  \includegraphics[width=4.5in]{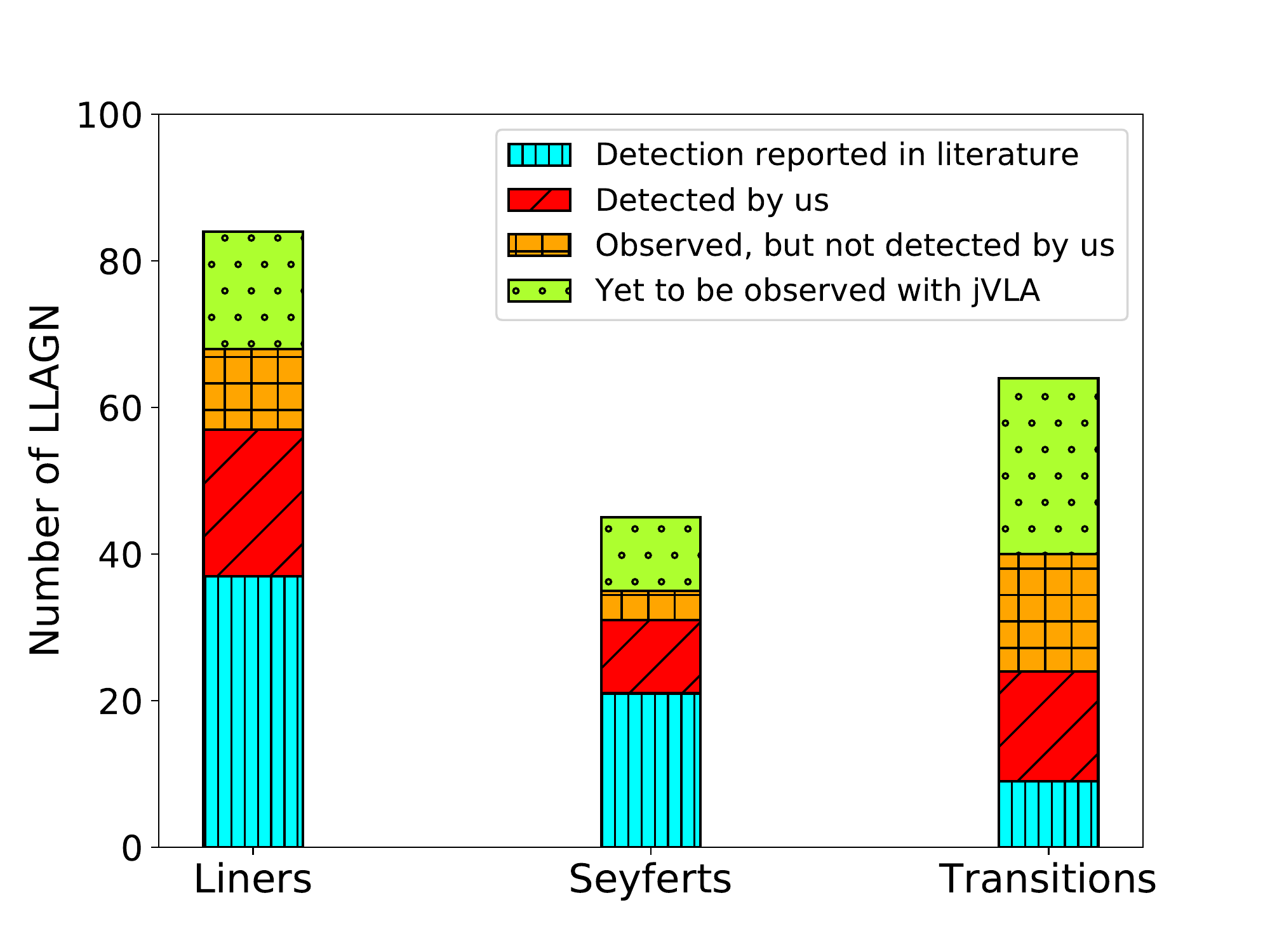}
}
\caption{Different types of LLAGN in the Palomar Sample (197 sources, classification taken from in \citet{h97}). The red (filled with vertical lines) bins represent the new detections reported in this paper, while the cyan (slanted lines) bins represent the earlier observations reported in the literature \citep[compiled in][]{n05}. The orange (crossed lines) bins depict the number of galaxies that were observed by us but not detected. Finally, the dotted region in green shows the number of galaxies yet to be observed with the highly sensitive VLA.}
\label{figdist}
\end{figure}

\begin{figure}[t]
\resizebox{\columnwidth}{!}
{
  \includegraphics[width=4.5in]{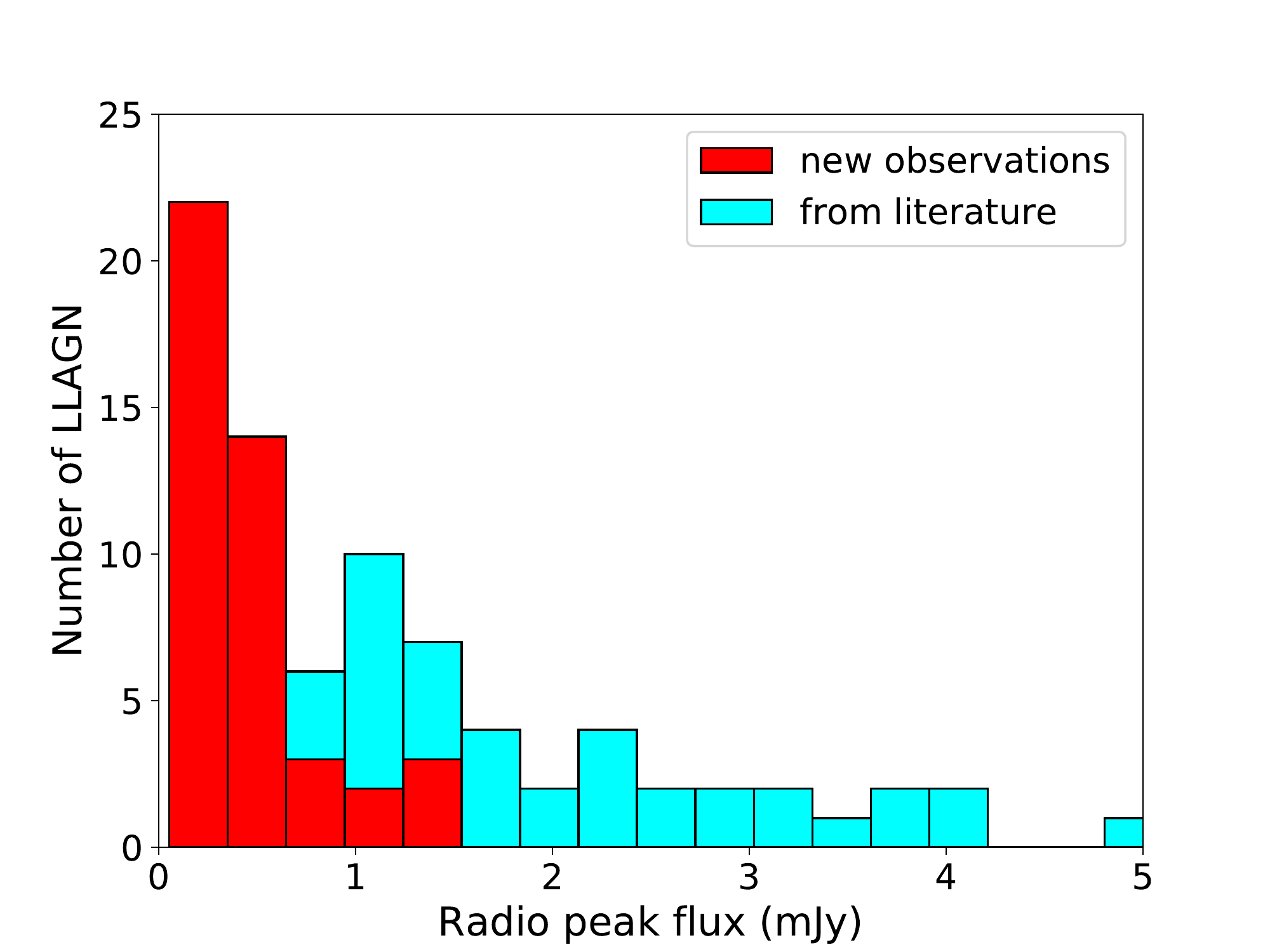}
}
\caption{Histogram of peak radio flux (in mJy) for the Palomar LLAGN sources. The red shaded bins represent the new detections reported in this paper, while the cyan dashed bins represent the earlier observations reported in the literature \citep[compiled in][]{n05}. We only show the fainter ($<$ 5mJy) systems here for clarity.}
\label{fighist}
\end{figure}

\begin{figure}[t]
\resizebox{\columnwidth}{!}
{
  \includegraphics[width=4.2in]{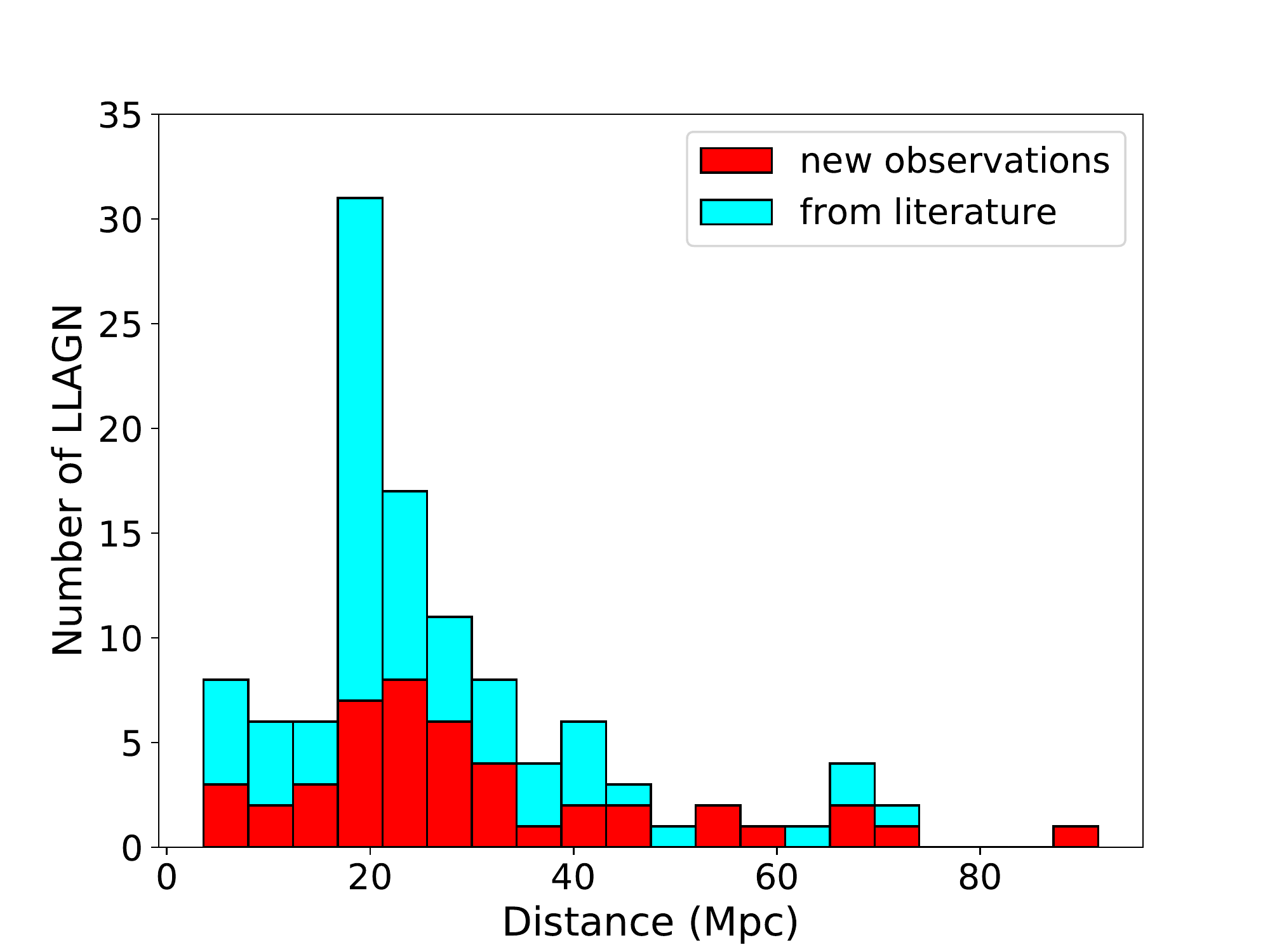}
}
\caption{Histograms of the distances for the sample of galaxies observed in this paper (red) and detected by previous studies in the literature (cyan). The peak in the histogram around D$\sim$17Mpc is caused by the members of the Virgo cluster.}
\label{figdist}
\end{figure}

A list of all the 45 sources detected in this survey, along with their radio properties (observed position, beamsize, peak flux, as well as error on the flux and rms in the map) are given in the Table 1. In Table 2, we list all the observations from this survey (all 76 sources) and compile the previous VLA detections of the Palomar LLAGN \citep[from][]{n05}. For both the tables, the columns are explained in the footnotes.

\begin{figure*}
\center
  \includegraphics[height=0.28\textwidth]{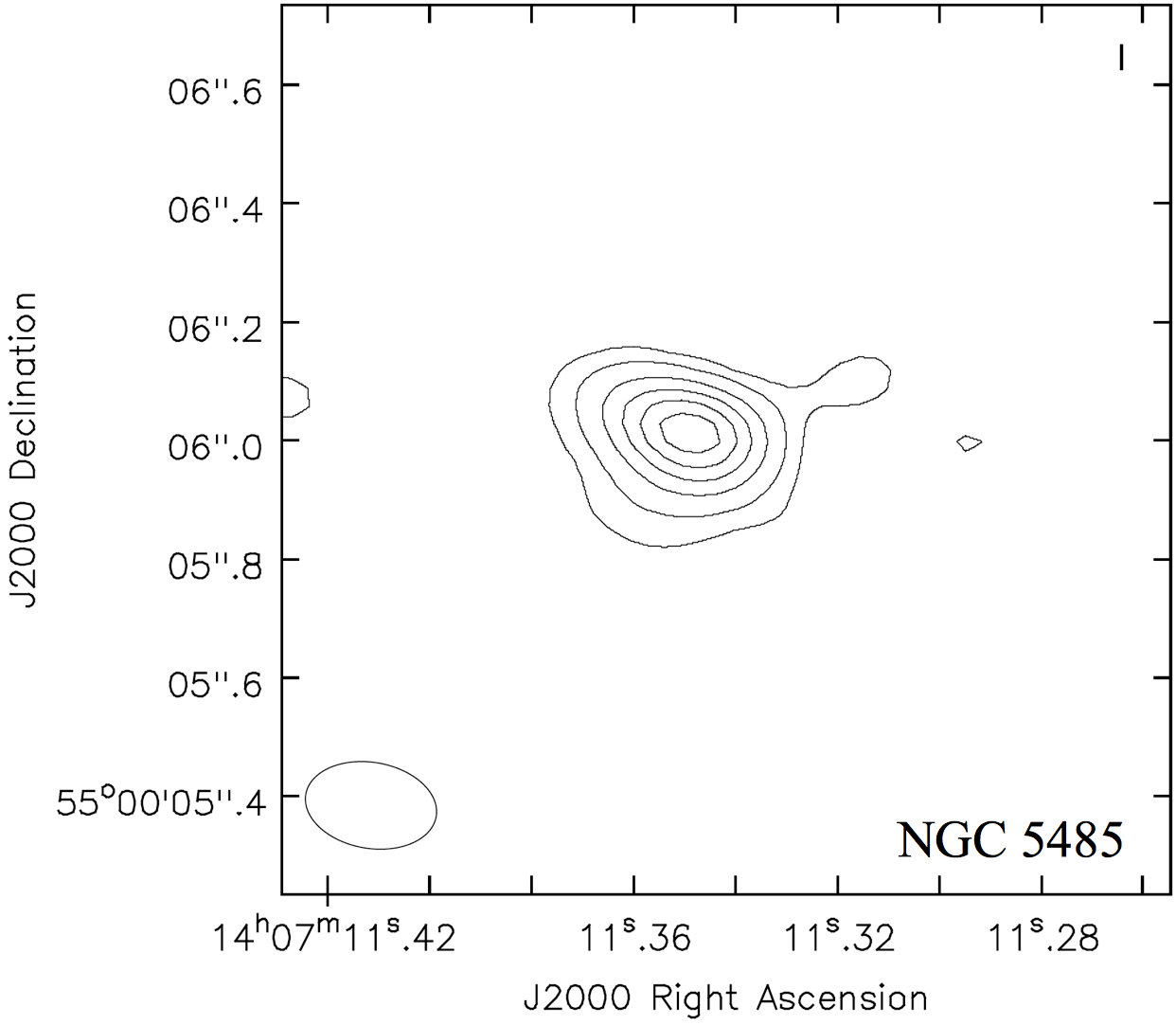}
  \includegraphics[height=0.28\textwidth]{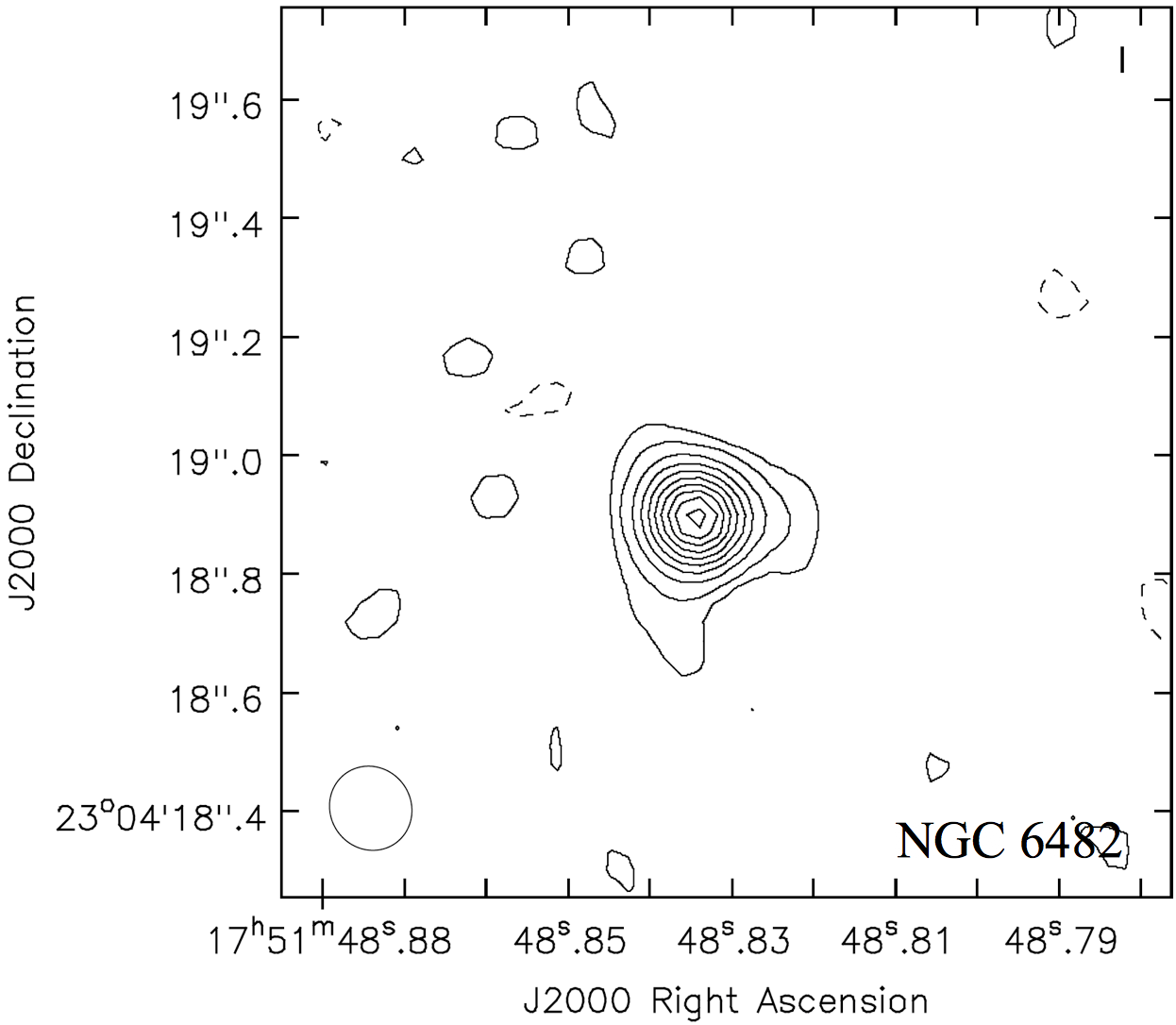}
  \includegraphics[height=0.28\textwidth]{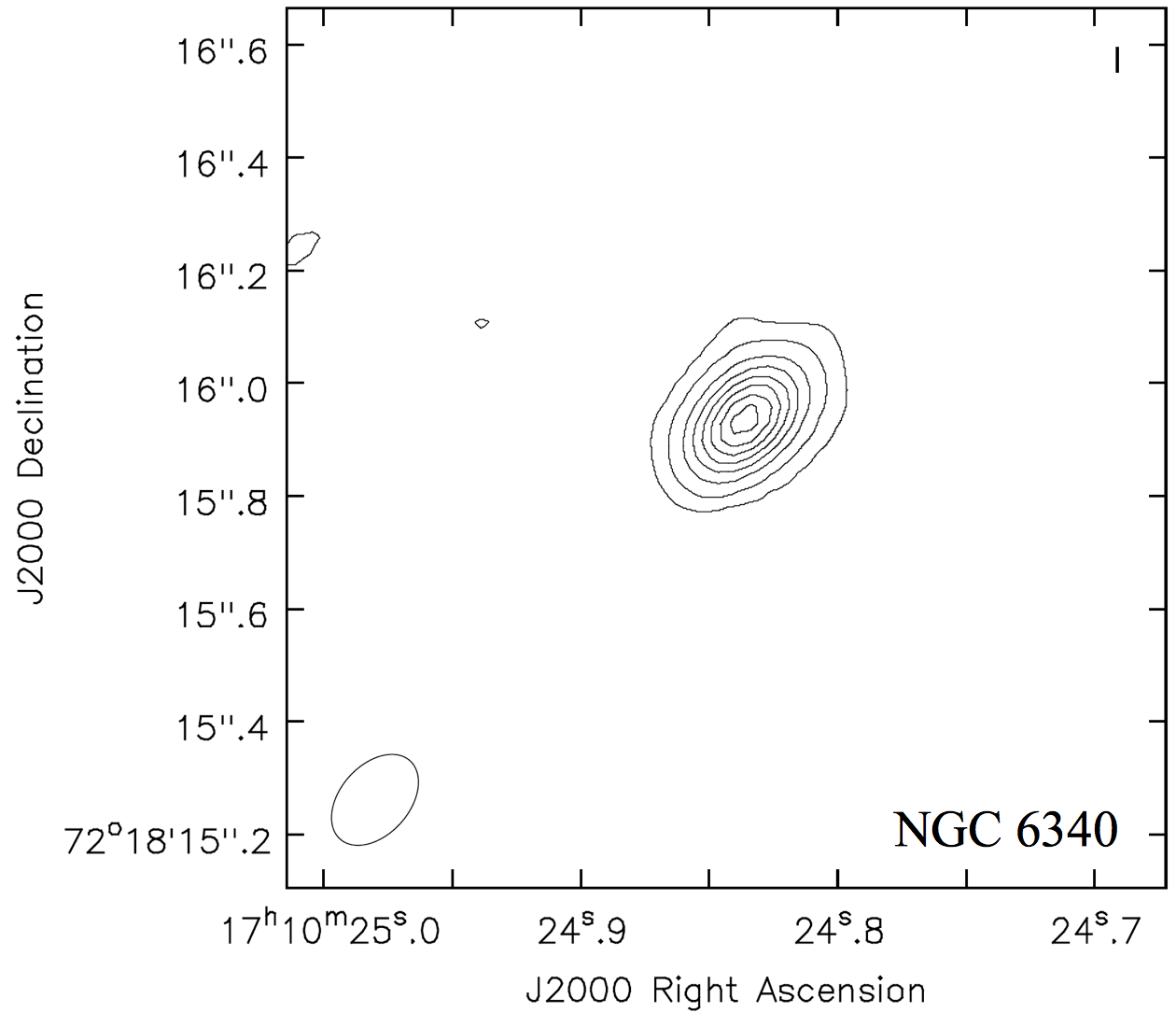}
  \includegraphics[height=0.28\textwidth]{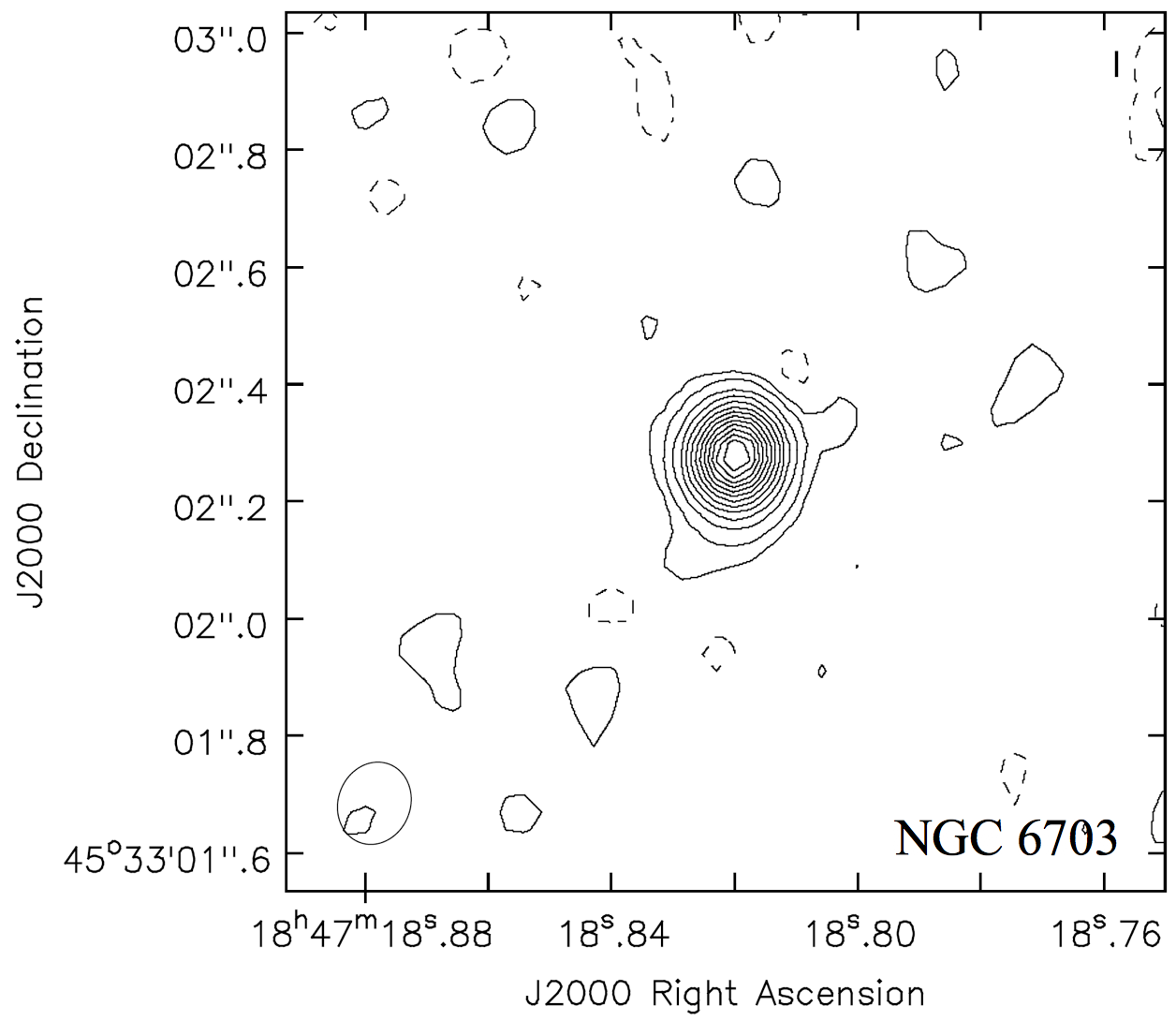}
  \includegraphics[height=0.28\textwidth]{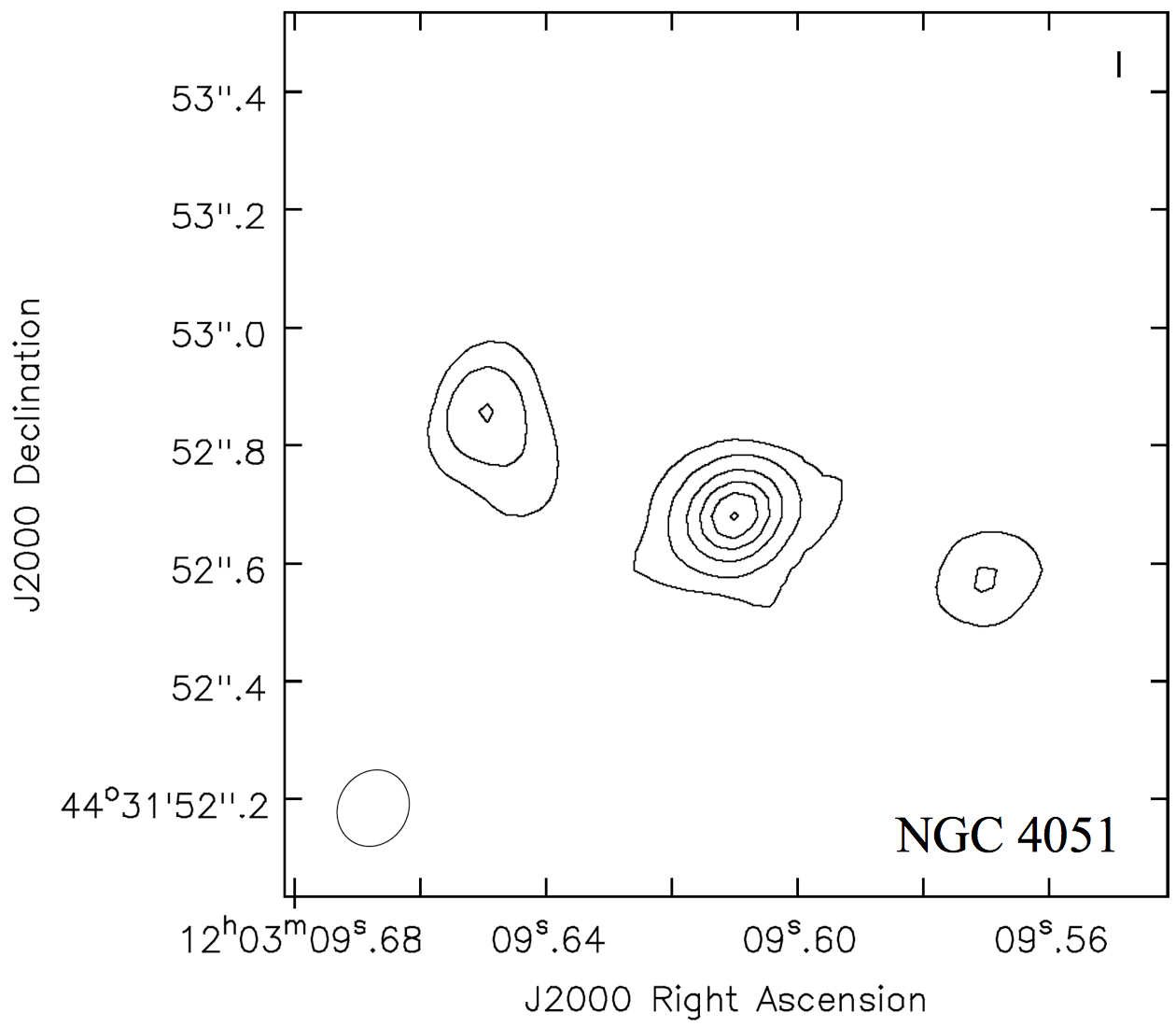}
  \includegraphics[height=0.28\textwidth]{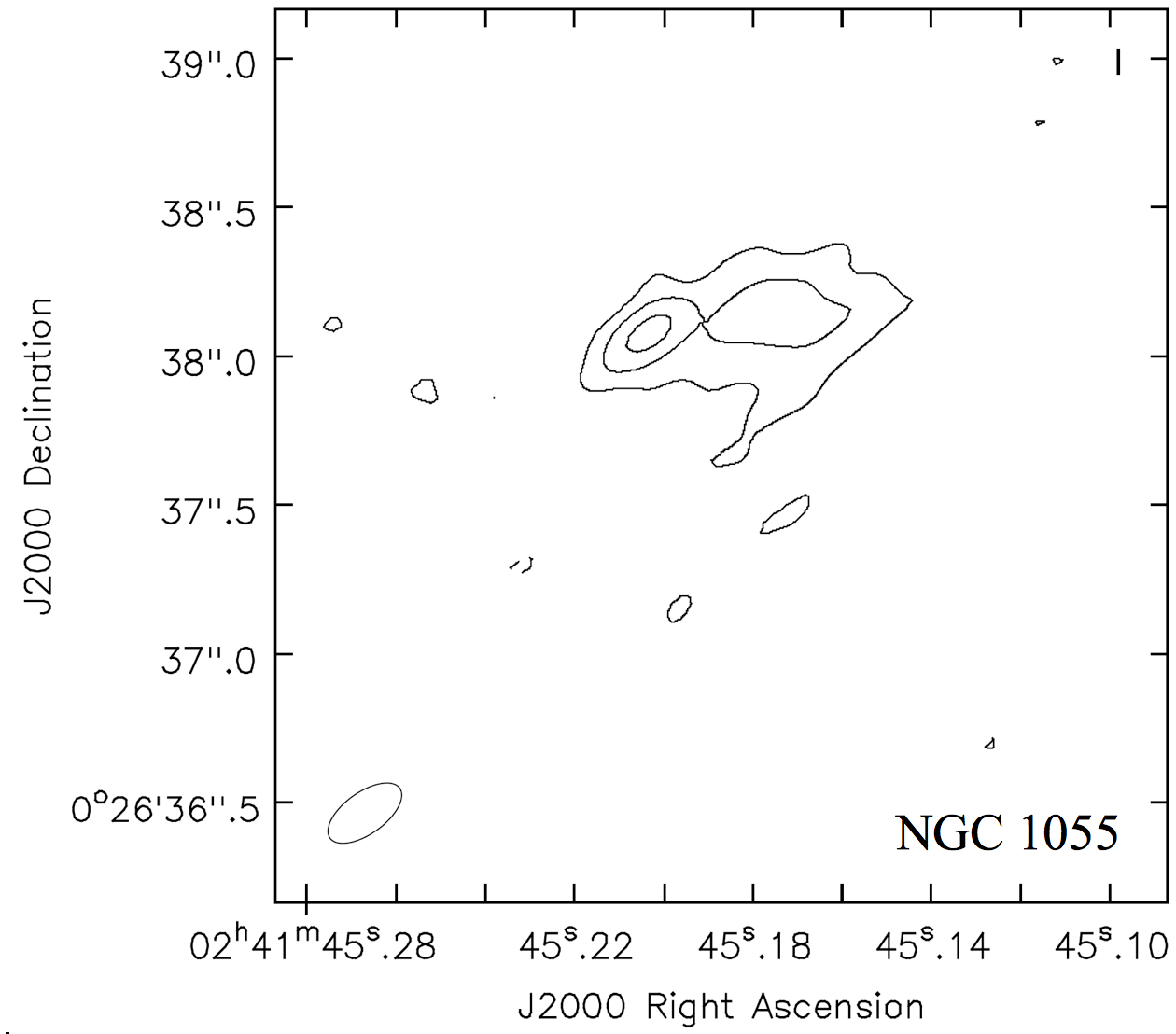}
  \includegraphics[height=0.28\textwidth]{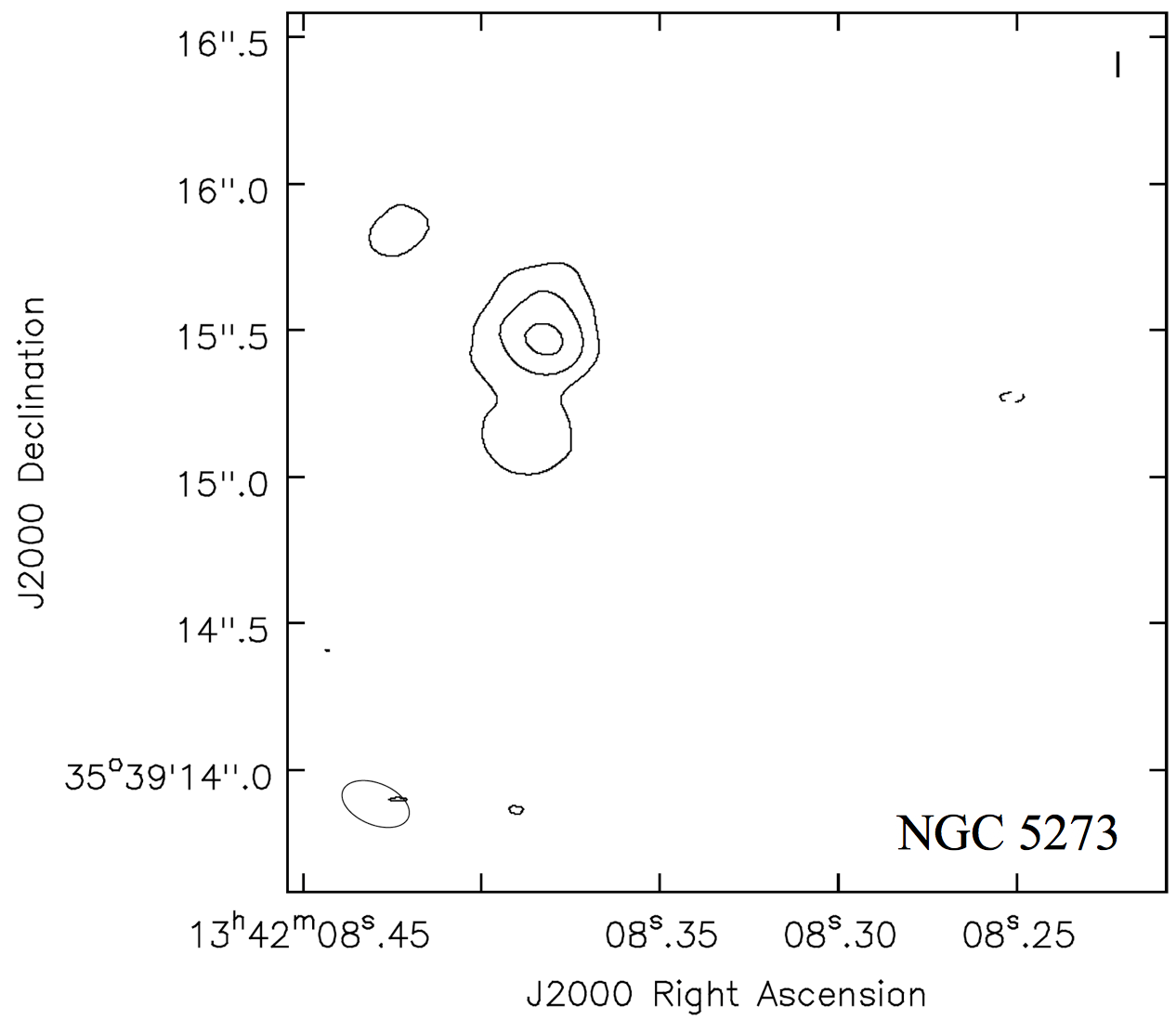}
  \includegraphics[height=0.28\textwidth]{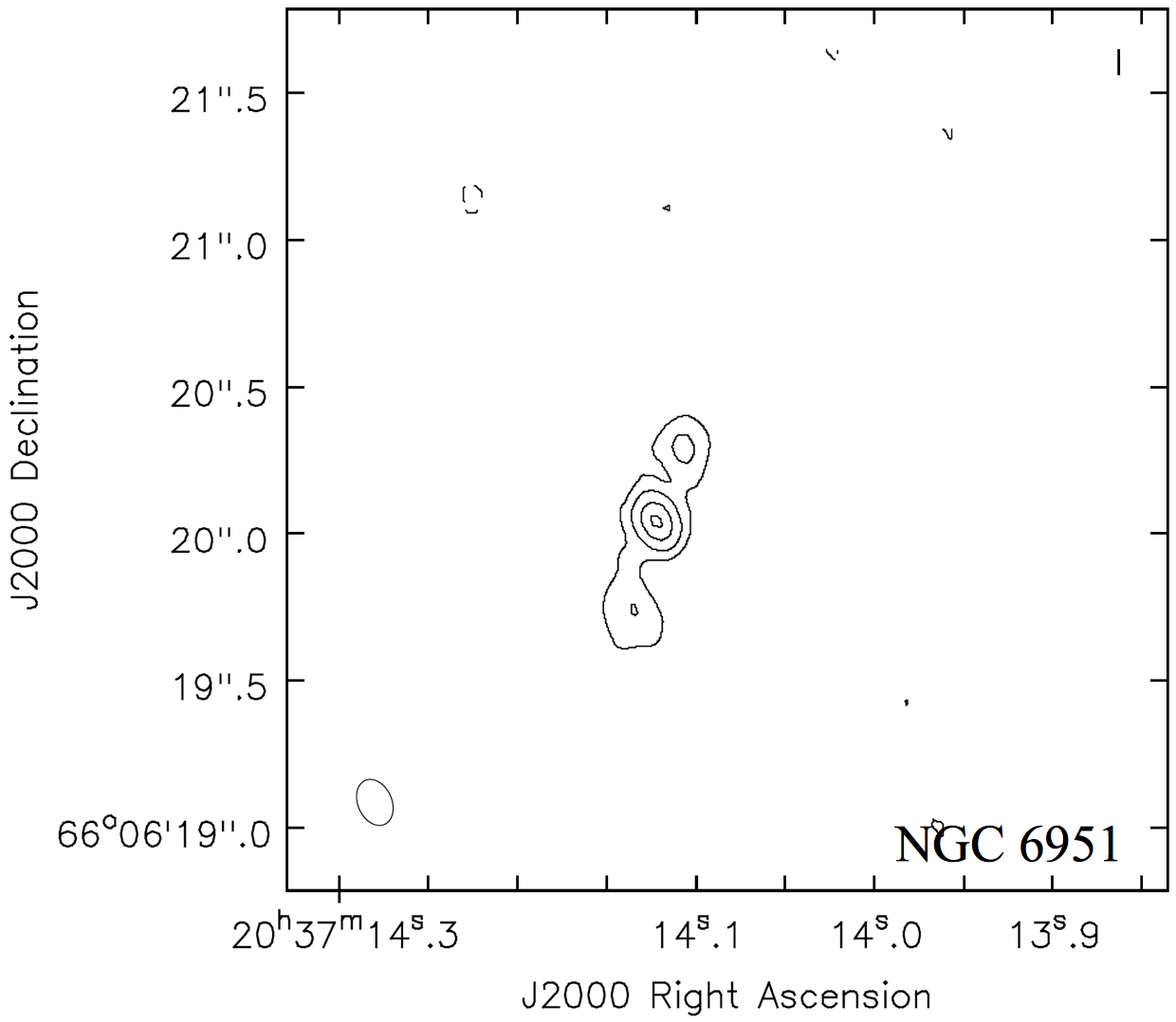}
  \includegraphics[height=0.28\textwidth]{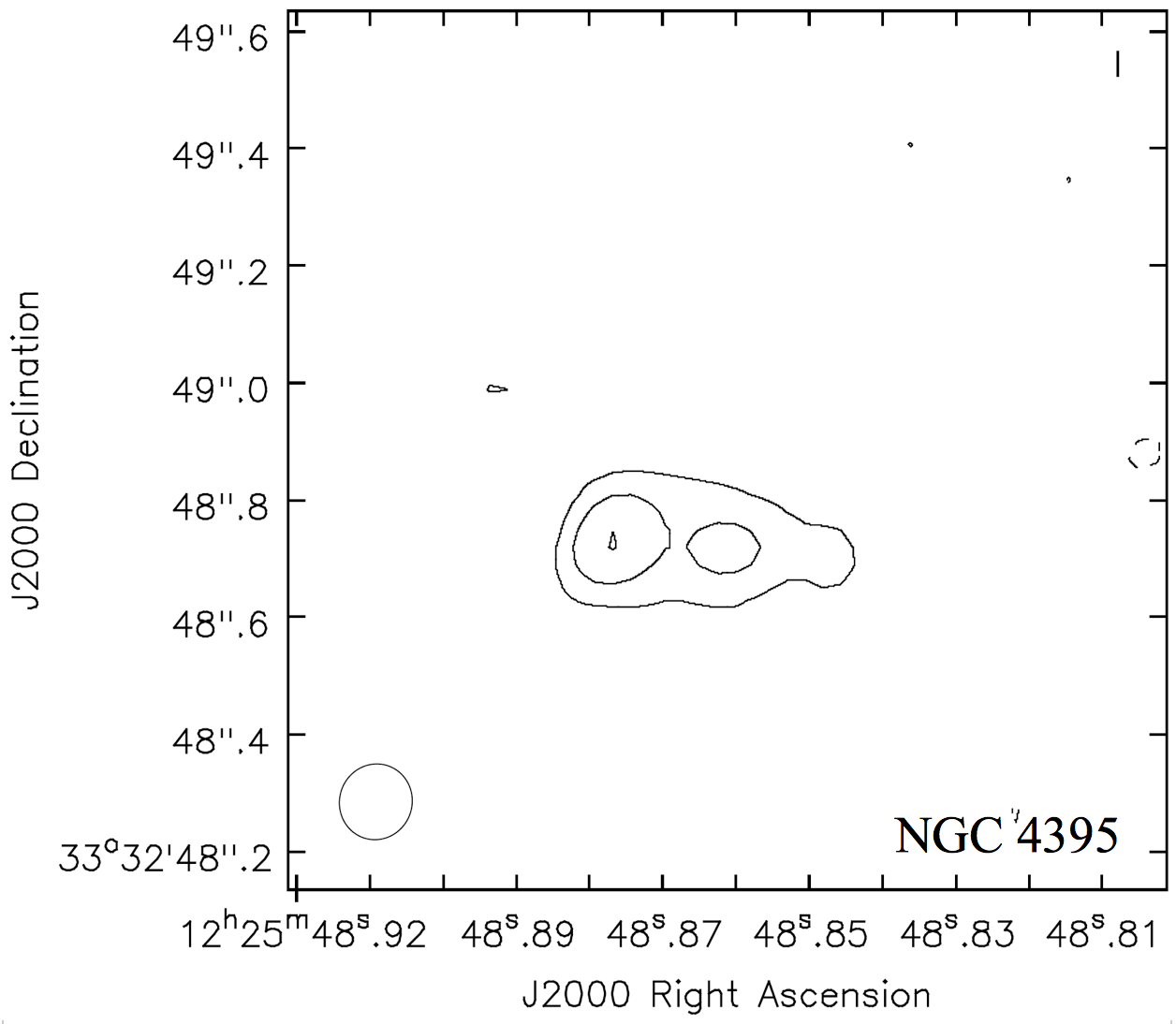}
\caption{VLA 15 GHz (2 cm) radio maps of four typical compact galaxies in the survey (NGC 5485, NGC 6482, NGC 6340, and NGC 6703 ). The remaining sources (NGC 4051, NGC 1055, NGC 5273, NGC 6951, and NGC 4395) plotted here are the only five detections with extended emission. The rms noise for individual maps are 13.1, 9.0, 10.2, 9.2, 12.3, 10.6, 12.0, 11.2, and 11.1 $\mu$Jy/beam, respectively. The lowest-level contours are at 3$\sigma$ rms noise for each source, while the subsequent contours are obtained at 3$\sigma$ rms noise multiplied by integer powers of 2.5. The dashed lines represent the negative contours (which show the -3$\sigma$ rms noise).}
\label{figcompact}
\end{figure*}

\subsection{Detection rates for different sources}

The complete sample of LLAGN and AGN in the Palomar survey has been sorted according to their nuclear activity by \citet{h97}. It comprises 45 Seyferts, 88 LINERs, and 64 transition (composite LINER + H II) nuclei. Of the complete sample, we observed 76 previously undetected sources. Our radio observations with the VLA have detected ten out of 14 (71$\%$) Seyferts, 20 out of 31 ($64\%$) LINERs, and 15 out of 31 ($48\%$) transition nuclei. The detection rates for different LLAGN types is shown in Fig 1. The complete detection rate is $\sim$60\% compared to the most complete survey until now (detection rate $\sim$35\%, Nagar et al 2005). The radio detections in this survey (45 sources) are compiled in Table 1, while the sample of LLAGN in the Palomar survey that were observed and reported in this paper (76 sources), as well as the previous radio detections obtained from the literature (67 sources) are compiled in Table 2. We exclude the sources that are reported as non-detections (at 15 GHz) in the literature.

\subsection{Measurement and errors on radio fluxes}

The peak radio flux densities of the detected radio nuclei and the errors on the flux measurements are given in Column 6 of Table 1. The radio flux distribution in the lowest flux range (0-5 mJy) is shown in Fig. 2. As seen in the figure, we are probing the very faint end of the LLAGN population, which was previously either undetected or under-populated in the literature.

The errors in the measured flux densities can arise due to many factors. The main sources of error are the accuracy of flux bootstrapping from a VLA flux calibrator ($\sim2.5\%$), variation of antenna gain with elevation (less than $1.5\%$ over the elevation range 35{\degr} to 65{\degr}, where our flux calibrators were observed), and atmospheric opacity ($\sim$0.02 at 2 cm for the VLA can result in a source having an observed flux that is $<$0.5$\%$ larger when observed at elevation 70{\degr}, compared to an observation at elevation 45{\degr}). The rms noise in the final maps is seen to be typically 11.5 $\mu$Jy/beam. Generally, the first factor, that is, the accuracy of the VLA flux calibrator, dominates the errors in the measured flux, which is typically $\sim$2.5$\%$ for the radio fluxes reported here \citep{n05}.

\subsection{Compactness of the detected radio emission}

The bulk of the detections reported here have no extended emission at 15 GHz. Only five of the 45 nuclei detected have evident extended emission - NGC 1055, NGC 5273, NGC 4395, NGC 4051, and NGC 6951. Three LLAGN in our sample, namely NGC 521, NGC 5850, and NGC 410, have hints of negligibly weak (less than 3$\sigma$) extended emission. The absence of extended emission in the rest of the nuclei can be explained by the use of high frequency (15 GHz, where extended emission is expected to be weak) and high resolution ($\sim$0.2$\arcsec$, which can resolve out most of the extended emission) observations. 

\subsection{Positional accuracy and distance measurements}

The positional accuracy of the detected radio nuclei are limited by the positional accuracy of the phase calibrators (typically 2-10 mas), and by the accuracy of the Gaussian fit to the source brightness distribution (depends on the signal to noise ratio of the detection). This brings the overall positional accuracy of the radio detections to $\sim$50 mas.

The distances for these sources (Column 3 of Table 2) are taken from the literature \citep{h09}. A histogram of the newly observed as well as previously detected LLAGN distances is shown in Fig. 3.

\subsection{Brightest source among the detections}

The brightest source that we have detected in our survey is NGC 660 (peak radio flux density of 62.8 mJy/beam). Interestingly, it was not detected by \cite{n05}, as the source was extremely faint when their observations were taken. In late 2012, a massive outburst emanating from NGC 660 was detected making the source many hundreds of times brighter than anything seen in the archive of radio images of NGC 660 \citep{argo2015}. High resolution images from the European VLBI Network (EVN) presented evidence of a high-speed jet of material leaving the vicinity of the black hole, and hence the onset of a new period of AGN activity in the core of NGC 660 \citep{argo2015}.\\

\section{Fundamental plane of black hole activity}
\label{thefp}

The fundamental plane of black hole activity is a non-linear empirical relation followed by both hard-state stellar-mass black holes and their supermassive analogs. It indicates a relationship between X-ray (or nuclear [OIII] emission line) luminosity, compact radio luminosity, and the mass of the black hole \citep{m03,f04,s15}. The radio luminosity probes the power of the jet \citep{bk79,fb99}, while the X-ray or the [OIII] line emission is expected to trace the accretion rate. It is however important to note that some studies have suggested that in a purely jet-dominated AGN system, most of the emission, including the X-rays, might come from the jet \citep[eg.,][etc.]{sara2001,extra2006}, but this is still an open question.

We use the new LLAGN detections reported in this paper to study the fundamental plane. For comparison with the stellar-mass XRBs, we use the sample of the best-studied XRBs in the low hard state, as the radio spectrum of a stellar-mass black hole in the low hard state is consistent with the spectra of a steady jet \citep{fbg04}. The sample used in this study consists of GX 339-4 \citep[88 quasi-simultaneous radio and X-ray observation from][]{c13}; V404 Cyg \citep[VLA radio and Chandra X-ray reported in][]{c08}; XTE J1118+480 \citep[from the compilation in][]{m03}, and A06200-00 \citep[simultaneous Chandra X- ray and VLA radio observation reported in][]{g06}.

\begin{figure}[t]
\resizebox{\columnwidth}{!}
{
  \includegraphics[width=4.8in]{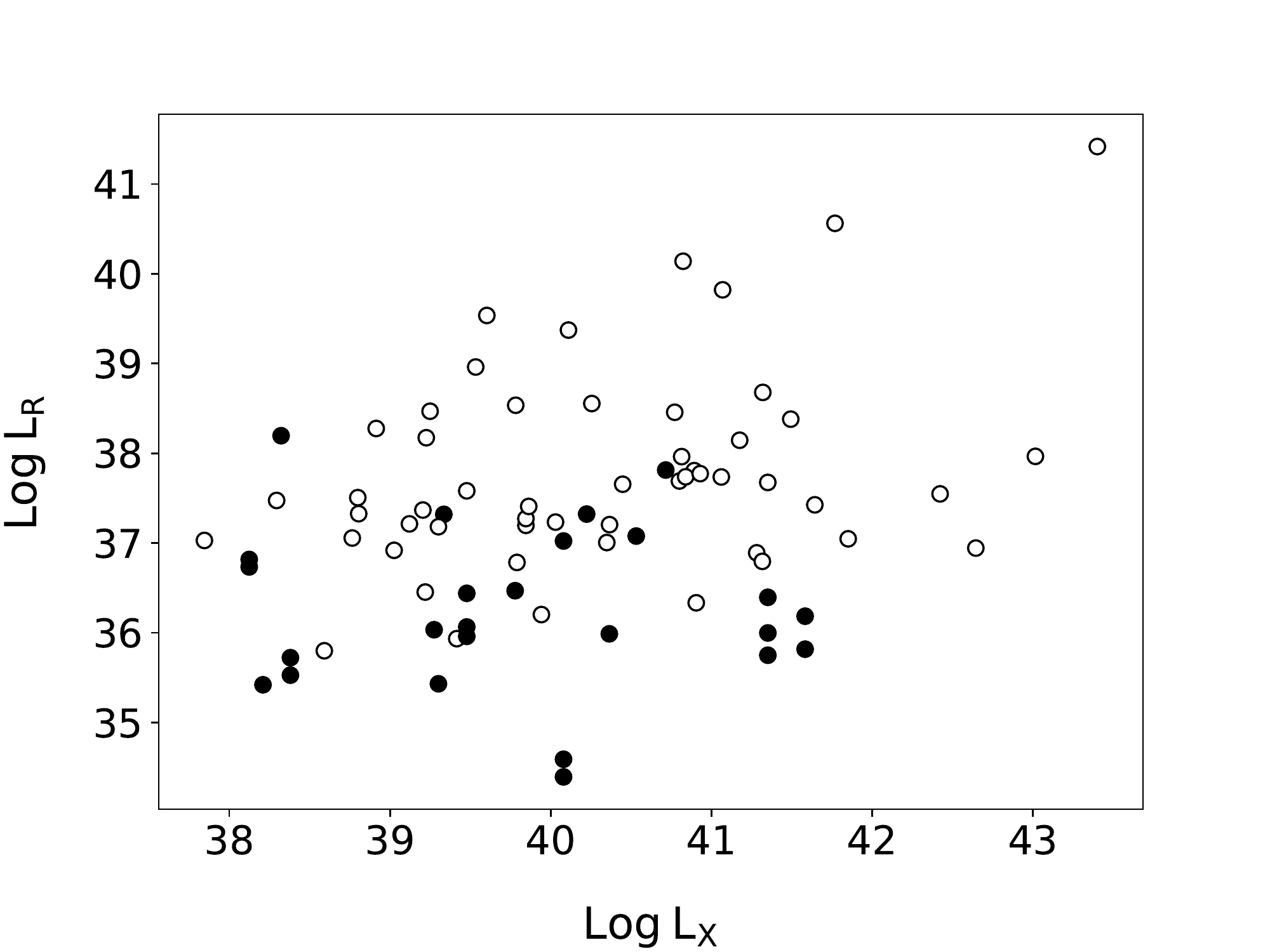}
}
\caption{Absence of a correlation between the radio luminosity obtained with the VLA versus the X-ray luminosity in the 2-10 keV range (Pearson correlation coefficient = 0.31, p-value = 0.0055). In our analysis, we include all the Palomar LLAGN that have been detected by the VLA at 15 GHz (see Table 3). While the new detections reported in this paper are shown with filled circles, the old detections are depicted by open circles.The relevant references to the X-ray data points are listed in the caption of Table 3. Luminosities are given in erg s$^{-1}$. }
\label{fig:rx}
\end{figure}

\begin{figure}
\includegraphics[width=3.4in]{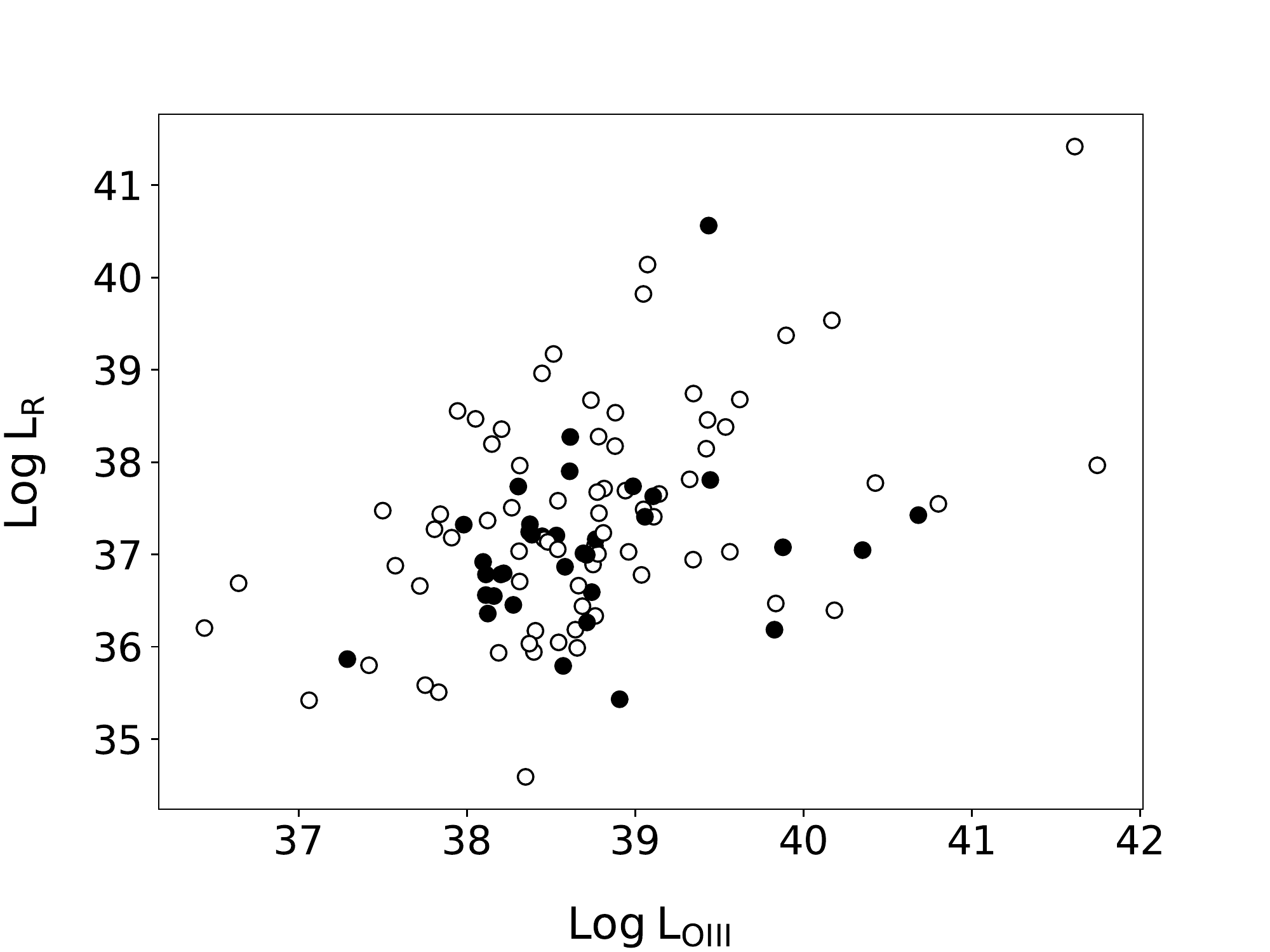}
\includegraphics[width=3.4in]{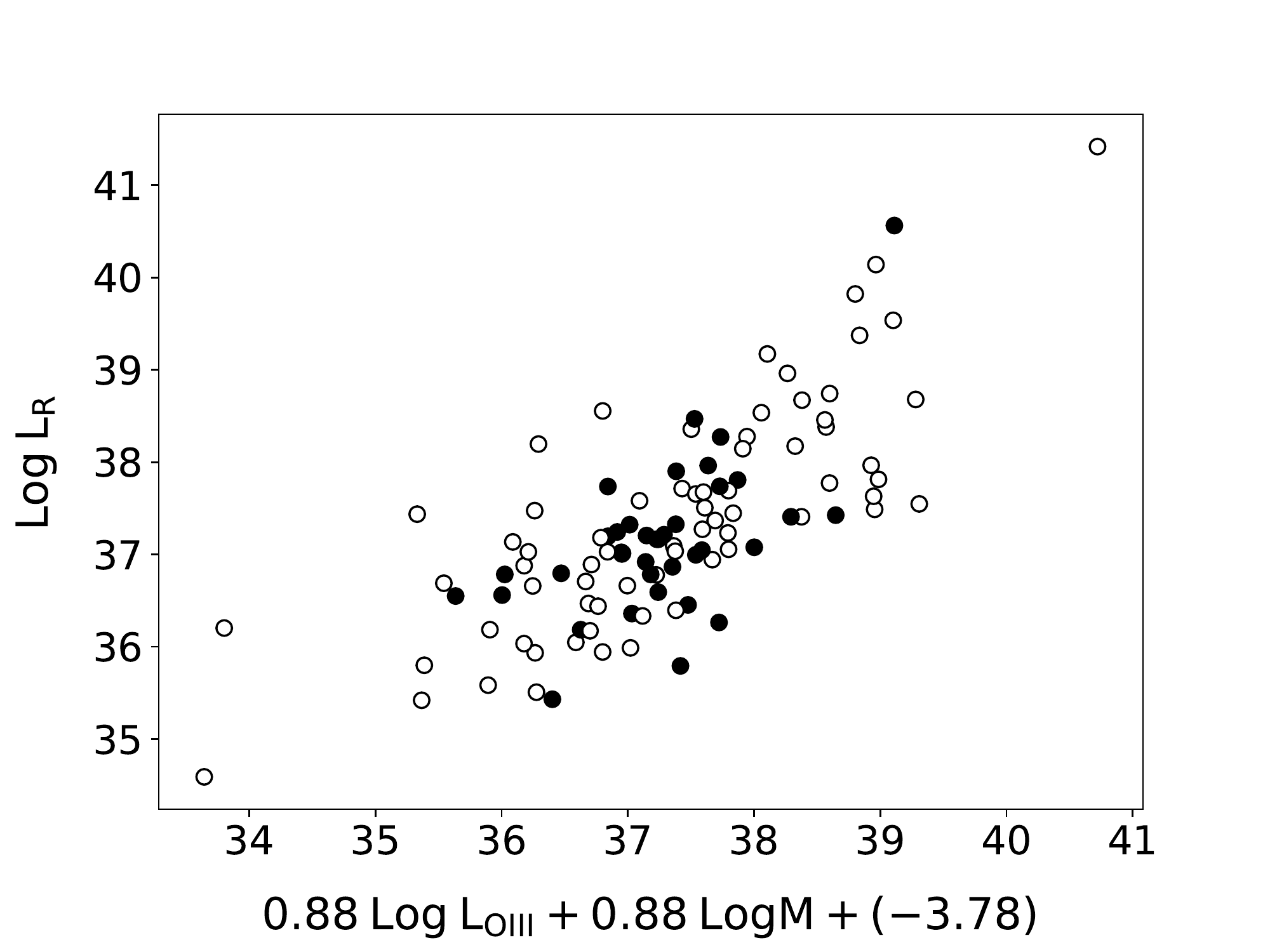}
\caption{Top figure shows the radio luminosity obtained with the VLA at 15 GHz and the nuclear [OIII] emission line luminosity (Pearson correlation coefficient = 0.43, p-value = 2.6 $\times 10^{-6}$). The bottom figure shows a tighter correlation after including the mass of the black hole as the third parameter. This is the edge-on view of the optical fundamental plane obtained by using only the LLAGN sample (Pearson correlation coefficient = 0.74, p-value = 4.2 $\times 10^{-20}$). While the new detections reported in this paper are shown with filled circles, the old detections are depicted by open circles. Luminosities are given in erg s$^{-1}$ while the masses are in units of solar mass.}
\label{ro}
\end{figure}

\begin{figure*}
 \center
  \includegraphics[width=0.89\textwidth]{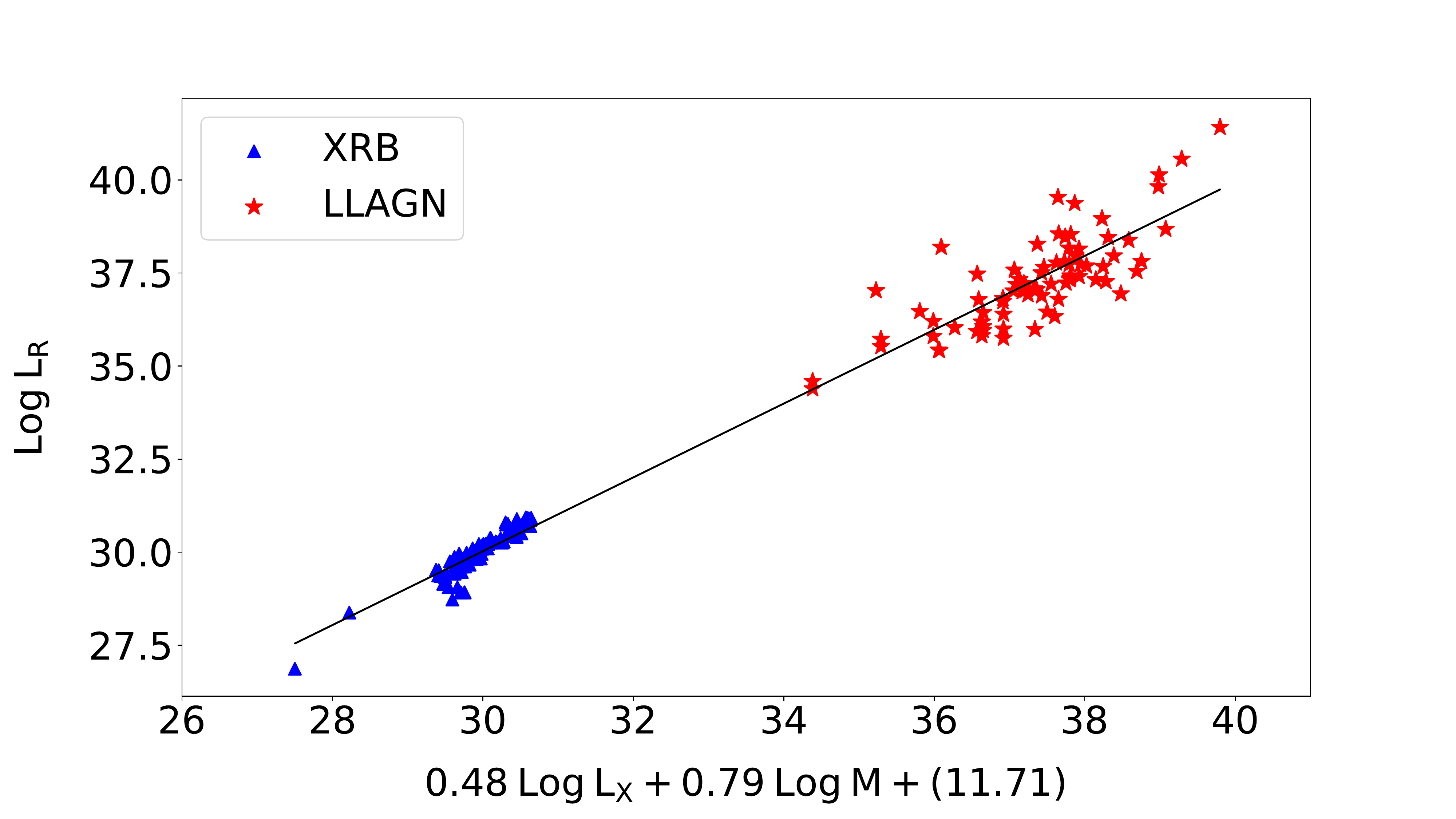}
  \includegraphics[width=0.89\textwidth]{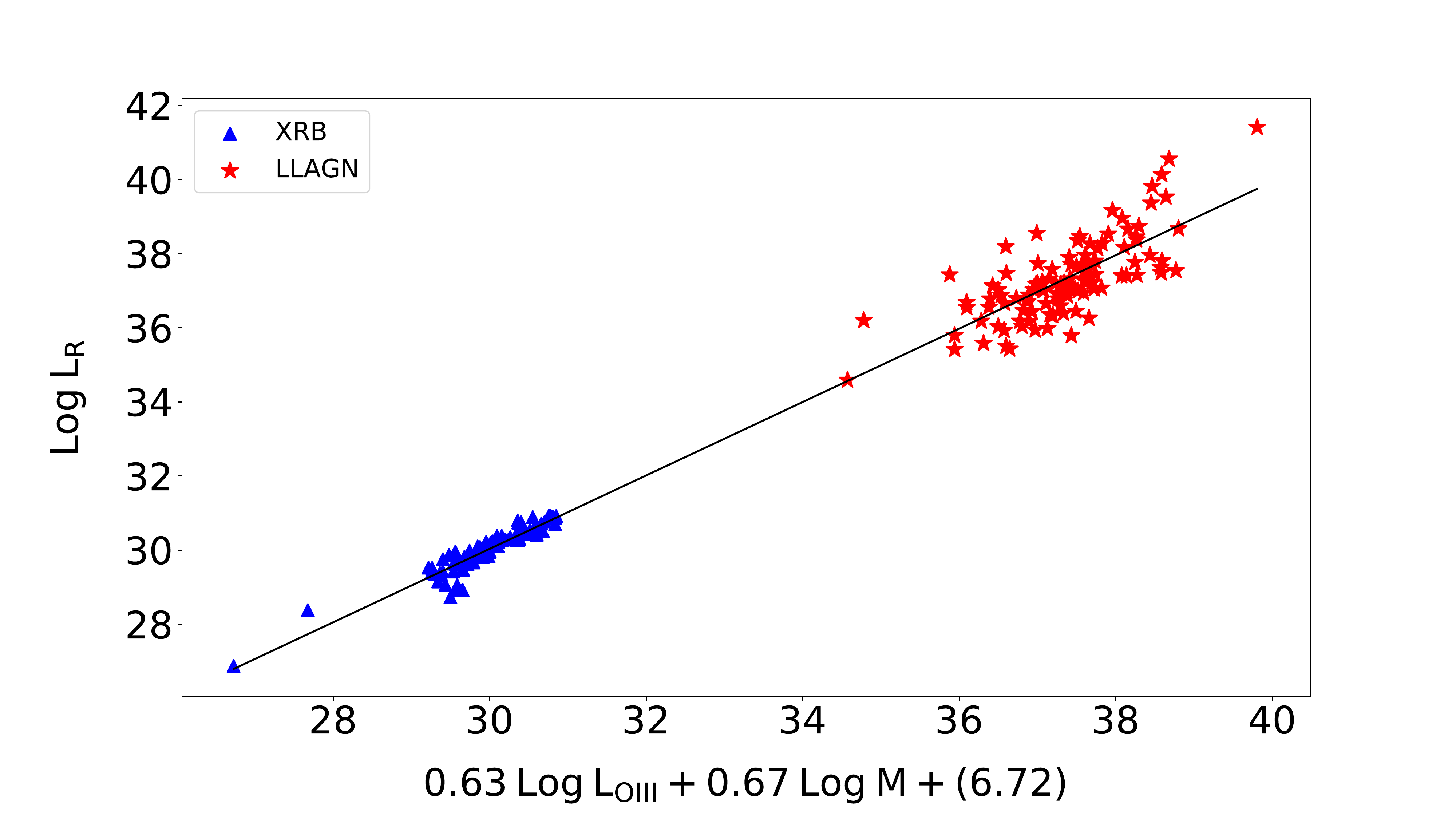}
  \caption{Projection of the fundamental and the optical fundamental plane of black hole activity. The LLAGN sample is shown in red stars, and the hard-state XRB sample is shown in blue triangles. The black line is the best-fit line of the complete sample. Luminosities are given in erg s$^{-1}$ while the masses are in units of solar mass. The relevant references to the X-ray data points are listed in the caption of Table 3.}
  \label{fp}
\end{figure*}

\subsection{Fundamental plane with the new sample}

In the previous studies of the fundamental plane, the radio fluxes of the supermassive black holes are only distributed over a bit more than two orders of magnitude, while the distances span a much wider range. A few studies have used radio fluxes extracted from the Faint Images of the Radio Sky at Twenty-Centimeters \citep[FIRST,][]{w97} survey, that observed with the VLA in B-configuration at 1.4 GHz to study the fundamental plane \citep[eg.,][etc.]{wang06,lww08}. Although it increases the number of AGN radio detections in the fundamental plane, it is uncertain whether all these radio flux densities trace the nuclear jet luminosity \citep{saikia18}.

With the new observations reported in this paper, we have probed the previously undetected LLAGN sample with high resolution ($\sim$0.2$\arcsec$) and higher sensitivity (typical rms $\sim$11.5$\mu$Jy/beam). For the construction of the fundamental plane, we have used the radio luminosity derived from the peak intensity. The new detections have more than doubled the number of LLAGN on the fundamental plane. In addition to the size of the sample, two other important advantages of the new observations used are the homogeneity of the sample (comprising only the low-luminosity AGN in the local universe) and the uniformity in the resolution and the sensitivity of the radio observations (VLA A-configuration, 15 GHz) used to refine the parameters of the fundamental plane.

The X-ray luminosity of the LLAGN sample at the 2-10 keV range is compiled from the previous literature (see Table 3 for references). We found hard X-ray luminosities for 78 of the total 121 LLAGN that were detected at 15 GHz. The black hole masses are estimated from the stellar velocity dispersions presented in \citet{h09}, with the empirical relation between black hole mass and central stellar velocity dispersions given in \citet{m11},

\begin{equation*}
\rm log \, \frac{M_{BH}}{M_{\odot}} = 8.29 + 5.12 \, log \, \frac{\sigma}{200 \, km \, s^{-1}}.
\end{equation*}

As shown in Fig \ref{fig:rx}, there is no statistically significant correlation observed between the nuclear radio and the X-ray luminosity (Pearson correlation coefficient = 0.31, p-value = 0.0055). In order to investigate the fundamental plane relation, we use the mass of the black hole as the third variable in our analysis. We perform a multivariate correlation linear regression analysis on these data using the modified chi-square estimator known as merit function to estimate the parameters for the plane. The best-fit plane relation obtained with only the supermassive black holes is log L$_\textrm{R}$ = $0.46 \, (\pm0.10$) log L$_\textrm{X}$ + $0.98 \, (\pm0.09$) log $\textrm{M}$, where luminosities are given in erg s$^{-1}$ while the masses are in units of solar mass. For the complete picture, we also include the best-studied hard-state XRBs, as described at the beginning of Sect~\ref{thefp}. As the luminosities of the XRBs were obtained from the literature at 5 GHz radio frequencies, we estimated the 5 GHz radio luminosities of the LLAGN sample from their 15 GHz detections assuming a flat radio spectrum. We find that the best-fit plane relation for the complete sample is log L$_\textrm{R}$ = $0.48\, (\pm0.04$) log L$_\textrm{X}$ + $0.79 \,(\pm0.03$) log $\textrm{M}$.

We perform a simple Kendall Tau rank correlation test to verify the significance of the plane. We get a Kendall Tau rank correlation coefficient value of $\tau$=0.86 (with the probability for null hypothesis as P$_{\rm null}$ $<$ 1 $\times$ 10$^{-20}$), hence showing that the plane relation obtained is statistically significant. In order to statistically check for spurious effects in luminosity-luminosity plots due to their common dependence on distance, we also perform a partial Kendall Tau correlation test with distance as the third variable. We get a Partial Kendall Tau coefficient value $\tau$ = 0.53, with a probability for null hypothesis as P$_{\rm null}$ = 1 $\times$ 10$^{-4}$, indicating that the correlation is weaker, but still statistically significant.

\subsection{Optical fundamental plane with the new sample}

The fundamental plane was recently discovered in the optical band using the nuclear [OIII] emission line luminosity instead of X-ray luminosity to trace the accretion power \citep{s15} using 39 LLAGN. \cite{lemm2018} used high angular resolution ($\sim$150 mas) observations of the Palomar LLAGN at 1.5 GHz with the eMERLIN array and found that, independent of their optical classes, the complete sample broadly follow the optical fundamental plane relation obtained in \cite{s15}. With the new detections reported in this paper, we have more than doubled the LLAGN sample used in the discovery paper.

In order to check for the parameters of the optical fundamental plane with our much larger and deeper sample, we use the [O III] line luminosities of the Palomar sample reported in \citep{h97}. We found no significant correlation between the observed radio luminosity at 15 GHz and the nuclear [OIII] line luminosity (Pearson correlation coefficient = 0.43, p-value =  2.6 $\times 10^{-6}$), but a more significant trend with much less scatter can be seen after including the mass of the black hole as the third parameter in the analysis (see Fig. \ref{ro}, Pearson correlation coefficient = 0.74, p-value = 4.2 $\times 10^{-20}$).

In order to recover the complete optical fundamental plane, we include the same sample of hard-state XRBs as described at the beginning of Sect \ref{thefp}. The [OIII] emission line luminosity of the LLAGN sample is converted to X-ray luminosity with the relation proposed by \cite{h05}, which states that the hard X-ray (3-20 keV) and [O III] line luminosities are well-correlated with the mean value for log (L$_{3-20 \rm keV}$/L$_{[\rm O~III]}$) as 2.15 dex. As the X-ray range used for our study is 2-10 keV instead of 3-20 keV, we use the modified relation of log (L$_{2-10 \rm keV}$/L$_{[\rm O~III]}$) = 1.81 dex, as described in \cite{s15}.

With this conversion factor, we estimate the luminosity of our LLAGN sample at 2-10 keV range from their observed [O III] line luminosity, and compare them with the XRBs in the hard state. We establish the optical fundamental plane best-fit relation as log L$_\textrm{R}$ = $0.63\,(\pm0.05)$ log L$_{[\rm O~III]}$ + $0.67\,(\pm0.03)$ log $\textrm{M}$, after including both the hard-state XRBs and the supermassive analogs, where luminosities are given in erg s$^{-1}$ while the masses are in units of solar mass.

We perform a Kendall Tau rank correlation test to check the significance of the plane. We find that the Kendall Tau rank correlation coefficient has a value of $\tau$ = 0.83 (with the probability for null hypothesis as P$_{\rm null}$ $<$ 1 $\times$ 10$^{-20}$), which shows that the plane correlation is statistically significant. The correlation is found to be real even after taking into account the large range of distances involved (Partial Kendall $\tau$ calculated with distance as the third variable, $\tau$ = 0.48, with P$_{\rm null}$ $<$ 1 $\times$ 10$^{-4}$).

\section{Radio luminosity function}
\label{therlf}

\begin{figure*}
 \center
 \includegraphics[height=0.6\textwidth]{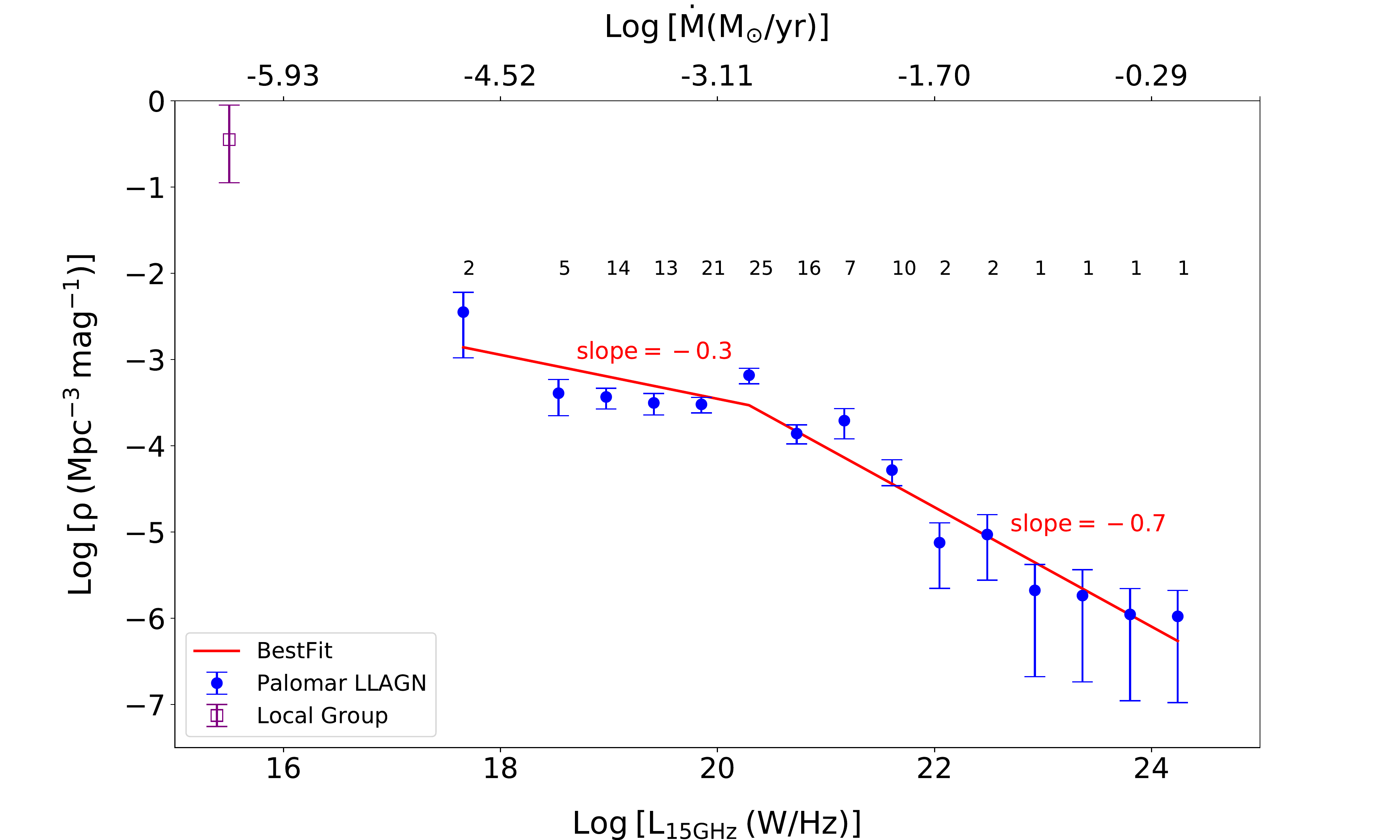}
  \caption{Radio luminosity function of the LLAGN in the Palomar sample. For each bin, the number of galaxies is shown at the top. Error-bars are assigned assuming Poisson statistics. For a rough comparison, we also plot the estimated position for galaxies in the local group \citep[from][shown in open square]{n05} in the 15 GHz nuclear RLF. The red line is the best-fit broken power law to the RLF. It has a slope of $\sim$-0.7 at the brightest end (complete sample, excluding the five lowest radio luminosity points), and $\sim$-0.3 for the fainter sources. The upper x-axis shows the implied logarithmic values of the mass accretion rate (in M$_{\odot}$/yr) within the context of a jet model, assuming a 10\% energy conversion efficiency (see Section 6.1 for details).}
  \label{rlf}
\end{figure*}

We construct the nuclear radio luminosity function (RLF) of the complete Palomar sample with the $\rm V/V_{max}$ method (Schmidt 1968). We include all the Palomar LLAGN that were detected either by us or by \cite{n05} at 15 GHz, and remove the non-detections and upper limits from our calculation. As our sample only has galaxies in the positive declinations, the survey area is restricted to 2$\pi$. The flux density limit of the radio survey is 40 $\mu$Jy, while the optical magnitude limit of the parent survey is 12.5 mag.

To construct the RLF, the sources are distributed over equal bins in log of radio power at 15 GHz. For each radio luminosity bin centered around a log L$_{\star}$, the space density or the differential RLF is calculated from the expression
\begin{equation*}
\rm \Phi \, (log \, L_{\star}) = \frac{4 \pi}{\sigma} \, \Sigma_{i=1}^{n(log L_{\star})} \frac{1}{V_{max(i)}},
\end{equation*}

where $\rm \frac{4 \pi}{\sigma}$ is the fraction of the sky covered by the survey, $\rm n (logL_{\star})$ is the number of the galaxies in the bin centered around $\rm L_{\star}$, and $\rm V_{max}$ is the maximum spherical volume in which the source could have been detected given both the magnitude limit of the parent survey and the radio flux limit, with the maximum volume being the smaller of the two. Error-bars are assigned assuming Poisson statistics. 

The RLF presented here is in rough agreement with the nuclear RLF of the Palomar sample reported in previous studies, for example, \citet[][for the Palomar Seyferts at 1.4 GHz and 15 GHz]{ulvho2001}, \citet[][for 35\% of the complete Palomar sample at 15 GHz]{n05}, and \citet[][for the complete sample using observations at multiple radio frequencies]{filho2006}. We also note that the constructed RLF at the highest luminosities is in good agreement with that of classical Seyferts.

\subsection{Broken power-law fit}

Compared to the Palomar nuclear RLF reported in \cite{n05}, we have heavily populated the number of LLAGN in the lower-luminosity bins, and have probed sources that are over one order of magnitude fainter (minimum $\rm log \, L^{peak}_{15GHz}$ of 17.21 W/Hz, compared to 18.62 W/Hz reported in \citealt{n05}). 

\cite{n05} had observed a tentative hint of the RLF flattening at the lower-luminosity end. With a much bigger sample of fainter sources detected in our survey, we can finally confirm the presence of a low-luminosity turnover in the nuclear RLF of LLAGN. We fit the RLF with a broken power law, with a best-fit power-law index of $\sim$-0.7 at the higher luminosities, and $\sim$-0.3 at lower luminosities. The slope at the fainter end of the RLF is comparable to the slope of -0.43 suggested by \cite{n05}. The break in the power law is seen around a peak radio luminosity of 2.0 $\times 10^{20}$ W/Hz.

In order to explore the physical meaning of the break, we convert the radio luminosities into an equivalent accretion rate. It has been shown by \cite{elmar06} that the flat spectrum radio luminosity of accreting objects can be used to trace the accretion rate of the black hole using the equation
 
 \begin{equation*}
\rm \dot M = 4 \times 10^{17} \, \Big(\frac{L_{Rad}}{10^{30} \, erg/s}\Big)^{12/17} \, \Big(\frac{0.1}{\eta} \, \frac{0.75}{f}\Big) \, \frac{g}{s},
\end{equation*}

where $\eta$ denotes the accretion disk efficiency and the factor `$f$' reflects what part of the total accretion rate is actually accreted and not ejected in the outflows. Following \cite{elmar06}, we have used $\eta$ = 0.1 and $f$ = 0.75 for our calculation. Using this equation, we find that the break in the nuclear RLF power law is seen when the mass accretion rate goes below 1.2$\times$10$^{-3}$ M$_{\odot}$/yr. For a rough comparison, X-ray and infrared emission from the Galactic center provides an upper limit of $\sim8\times10^{-5}$ M$_{\odot}$/yr on the mass accretion rate in Sgr A$^{\star}$ \citep{q99}.

We can also estimate the expected Eddington ratio at the turnover radio luminosity, by using an estimated black hole mass and the observed luminosity or mass accretion rate. The expected Eddington ratio, or the dimensionless mass accretion rate, is given as $\rm \dot m_{Edd} = \dot M/\dot M_{Edd}$, where $\rm \dot M_{Edd} = L_{Edd}/\eta c^2$ and $\rm L_{Edd} =1.2\times 10^{31} \frac {M}{M_{\odot}}$ W. Assuming a black hole mass of 10$^9 M_{\odot}$, we obtain a critical Eddington ratio of 5.1 $ \times 10^{-5}$ below which the RLF turns over.  The sources have very low inferred Eddington ratios, which require either a very low mass accretion rate, a radiatively inefficient accretion mechanism or both. 

\subsection{Comparison with the local group}

It is interesting to note that the local group falls between the extrapolation of the RLF for brighter sources (with a slope $\sim$-0.7), and the extrapolation of the flatter RLF after the turn-over for the fainter sources (with a slope $\sim$-0.3). The local group RLF is taken from \cite{n05}. They used a simple V/Vmax test for the Galaxy and M 31, assuming a detection limit of 30 $\mu$Jy, to estimate the position of the local group in the nuclear RLF of local universe. 

The discrepancy of the position of the local group on the RLF, in comparison with other LLAGN and the brighter seyferts, is not clearly understood. The local group is located much higher than the extrapolation of the RLF of the fainter local sources. One possible reason could be presence of slightly different physics near to the Galactic center, as it mainly feeds on stellar winds, making the accreting environment different from the active galaxies.

%Maybe our LG is special, since we are in a special time). It is still a bit strange that their number is so high. Hence, maybe that is some kind of a floor level of accretion that is rock bottom and that?s where galaxies pile up

\section{Conclusion}
\label{secconclusion}

In this paper, we present the results of a sub-arcsec resolution radio survey of 76 LLAGN and AGN in the Palomar sample of nearby bright northern galaxies that were previously not detected with the VLA at 15 GHz. This sample concentrates on the active side of the Palomar sample, as HII galaxies and absorption line galaxies are not included in our analysis. With a total time of approximately four minutes per source, we have a typical rms noise level of 11.5 $\mu$Jy/beam compared with the previous best in literature, 0.3 mJy \citep{n05}. We detect 60 \% (45 of 76) of this previously non-detected LLAGN sample for the first time at 15 GHz, comprising 20 low-ionization nuclear emission-line region (LINER) nuclei, ten low-luminosity Seyfert galaxies, and 15 transition objects.

We use the new observations of fainter LLAGN to populate and refine the fundamental plane of black hole activity. We find that the fundamental plane relation is significant even after removing the common effect of distances on the two axes of the plane. The best-fit plane relations were found to be log L$_\textrm{R}$ = $0.48\,(\pm0.04$) log L$_\textrm{X}$ + $0.79\,(\pm0.03$) log $\textrm{M}$ and log L$_\textrm{R}$ = $0.63\,(\pm 0.05)$ log L$_{[\rm O~III]}$ + $0.67\,(\pm 0.03)$ log $\textrm{M}$.

We also construct the nuclear RLF of the LLAGN sample in the local universe, using our uniform, high-resolution and high-frequency radio detections of the Palomar sample. We have probed very weak sources that are an order of magnitude fainter than previously detected at nuclear radio luminosities, and have provided significant evidence of a LLAGN nuclear RLF turnover at lower luminosities. We fit a broken power law to the 15 GHz RLF of the Palomar active galaxies, and find that the power-law break occurs at a critical mass accretion rate of 1.2$\times$10$^{-3}$ M$_{\odot}$/yr (which translates to an Eddington ratio $\rm \dot m_{Edd} \sim 5.1 \times 10^{-5}$, assuming a black hole mass of 10$^9 M_{\odot}$). We also note that the local group stands out and falls closer to the extrapolation of the higher-luminosity sources. The constructed RLF agrees with the classical Seyferts at higher luminosities.

\section*{ACKNOWLEDGEMENTS}
We acknowledge funding from an NWO VIDI grant under number 639.042.218. HF acknowledges funding from the ERC Synergy Grant `BlackHoleCam-Imaging the Event Horizon of Black Holes' (Grant 610058). IMcH acknowledges support from a Royal Society Leverhulme Trust Senior Research Fellowship LT160006. RB and IMcH also acknowledge support from STFC grant ST/M001326/1. DRW acknowledges support from a University of Southampton Mayflower Studentship. This publication uses radio observations carried out using the National Radio Astronomy Observatory facility Karl G. Jansky Very Large Array (VLA). The National Radio Astronomy Observatory is a facility of the National Science Foundation operated under cooperative agreement by Associated Universities, Inc. 

%\begin{figure*}
%\center
  %\includegraphics[height=0.39\textwidth]{AAA.png}
  %\includegraphics[height=0.39\textwidth]{BBB.png}
  %\includegraphics[height=0.39\textwidth]{CCC.png}
  %\includegraphics[height=0.39\textwidth]{DDD.png}
  %\includegraphics[height=0.39\textwidth]{EEE.png}
%\caption{15 GHz (2 cm) VLA maps of (left to right) NGC 2273, NGC 4589, NGC 5353, NGC 5363, and NGC 7626. The contours are integer powers of 3.5, multiplied by the 3sigma noise level.}
%\label{figcompact}
%\end{figure*}

%\begin{figure*}
%\center
  %\includegraphics[height=0.39\textwidth]{CA.png}
  %\includegraphics[height=0.39\textwidth]{CB.png}
  %\includegraphics[height=0.39\textwidth]{CD.png}
  %\includegraphics[height=0.39\textwidth]{CC.png}
%\caption{15 GHz (2 cm) VLA maps of (left to right) NGC 2273, NGC 4589, NGC 5353, NGC 5363, and NGC 7626. The contours are integer powers of 3.5, multiplied by the 3sigma noise level.}
%\label{figcompact}
%\end{figure*}

%

\nopagebreak

\onecolumn
\begin{longtable}{lllllll}
\caption{Radio properties of the 45 detected sources.}\\ 
\hline
\hline
Name   & $\rm RA$    & $\rm Dec$ & Beam  & BPA & $\rm S^{peak}_{15 GHz}$  & RMS  \\
        &      (h:m:s)  & (d:m:s)       & (arcsec)    & (deg)   & (mJy/beam)       &   ($\mu$Jy/beam)    \\
            (1)    & (2)  & (3)        & (4)       & (5)       & (6)   & (7)\\
        \hline
        \\
        \endhead

NGC410 & 01:10:58.91  & 33:09:07.1 & 0.18$\times$0.13 & -82.6 & 0.72$\pm$0.01 	 & 11.1    \\
NGC488  & 01:21:46.80  & 05:15:24.6 & 0.22$\times$0.14 & -58.3 & 0.14$\pm$0.01  	& 10.7  \\
NGC521 & 01:24:33.77  & 01:43:52.9  & 0.23$\times$0.14 & -53.3 & 0.34$\pm$0.01	& 10.6       \\
NGC660  & 01:43:02.32  & 13:38:44.9  & 0.21$\times$0.14 &  -66.2 & 62.80$\pm$0.12	 & 9.0      \\
NGC777  & 02:00:14.91  & 31:25:45.9  & 0.22$\times$0.13  & -76.4 &  0.53$\pm$0.01 	& 10.7    \\
NGC841  & 02:11:17.37  & 37:29:49.7  & 0.21$\times$0.13  & -79.9 & 0.06$\pm$0.01    & 7.8   \\
NGC1055 & 02:41:45.20  & 00:26:38.1 & 0.29$\times$0.14 & -54.4 & 0.18$\pm$0.01    & 10.6      \\
NGC1169 &  03:03:34.74  & 46:23:11.1  & 0.23$\times$0.13 &  -81.0 & 0.36$\pm$0.01	 &  9.8      \\
NGC1961 & 05:42:04.65  & 69:22:42.4 & 0.26$\times$0.13 &  72.8 & 0.50$\pm$0.01	& 9.5      \\
NGC2681 & 08:53:32.72  & 51:18:49.2 & 0.17$\times$0.14 & -87.2 & 0.34$\pm$0.01	& 9.6     \\
NGC2683 &   08:52:41.30  & 33:25:18.7 & 0.19$\times$0.13 & 65.6 & 0.45$\pm$0.01	& 10.8      \\
NGC2832 &  09:19:46.85  & 33:44:59.0 & 0.17$\times$0.13 & 67.5 & 0.20$\pm$0.01	 & 9.2    \\
NGC2859 &  09:24:18.53  & 34:30:48.6 & 0.17$\times$0.13 & 65.2 & 0.05$\pm$0.01    & 8.9    \\
NGC2985 &  09:50:22.18  & 72:16:44.2 & 0.18$\times$0.14 &  -46.0 & 1.14$\pm$0.01	& 12.3      \\
NGC3642 & 11:22:17.90   & 59:04:28.3  & 0.15$\times$0.12 &  15.0 &  0.78$\pm$0.01	& 10.5     \\
NGC3735 & 11:35:57.22  & 70:32:07.8  & 0.16$\times$0.12 &  9.4 & 0.39$\pm$0.01	& 9.7     \\
NGC3898 & 11:49:15.24  & 56:05:04.3 & 0.14$\times$0.12 &  20.2 & 0.21$\pm$0.01	& 9.0    \\
NGC3982 & 11:56:28.14  & 55:07:30.9 & 0.14$\times$0.12 &  22.0  & 0.56$\pm$	0.01 & 10.2     \\
NGC3992 &  11:57:35.96  & 53:22:29.0  & 0.14$\times$0.12 &  25.5 &  0.17$\pm$0.01	& 9.9      \\
NGC4036 & 12:01:26.75  & 61:53:44.6 & 0.15$\times$0.12 &  10.6  & 0.60$\pm$	0.01 & 11.6     \\
NGC4051 & 12:03:09.61  & 44:31:52.7  & 0.13$\times$0.12 &  -31.6 & 0.48$\pm$0.01   & 12.3      \\
NGC4111 & 12:07:03.13  & 43:03:56.3 & 0.13$\times$0.12 &  -31.4 & 0.19$\pm$	0.01 & 12.1     \\
NGC4145 &  12:10:01.56  & 39:53:00.9 & 0.14$\times$0.12 &  -32.7 &  0.09$\pm$0.01   & 9.2      \\
NGC4346 & 12:23:27.96  & 46:59:37.6 & 0.14$\times$0.12 &  -32.6 &  0.06$\pm$0.01   & 8.3      \\
NGC4395 & 12:25:48.88  & 33:32:48.7  & 0.13$\times$0.12 &  -17.9  &  0.17$\pm$0.01     & 11.1     \\
NGC4750 & 12:50:07.32  & 72:52:28.6  & 0.17$\times$0.12 &  15.6 & 0.86$\pm$0.01	& 12.5    \\
NGC5194 & 13:29:52.71  & 47:11:42.8 & 0.26$\times$0.15 &  70.5 & 0.25$\pm$0.01	& 14.2  \\
NGC5195 & 13:29:59.53  & 47:15:58.3 &  0.27$\times$0.14 &  69.5 & 0.25$\pm$	0.01 & 14.3   \\
NGC5273 & 13:42:08.38  & 35:39:15.5 & 0.24$\times$0.14 &  67.5  &  0.19$\pm$0.01   & 12.0 \\
NGC5395 & 13:58:37.96  & 37:25:28.2 & 0.23$\times$0.13 &  67.2 & 0.11$\pm$0.01	& 12.3   \\
NGC5448 & 14:02:50.05  & 49:10:21.2 & 0.25$\times$0.14 &  71.8 & 0.06$\pm$0.01   & 12.3    \\
NGC5485 & 14:07:11.35  & 55:00:06.0  &  0.22$\times$0.15 &  80.4  & 0.56$\pm$0.01	& 13.1   \\
NGC5566 & 14:20:19.89  & 03:56:01.4 & 0.28$\times$0.13 &  48.6 & 1.40$\pm$0.01  & 29.3	     \\
NGC5631 & 14:26:33.29  & 56:34:57.4 & 0.17$\times$0.14 &  -86.1 &  0.64$\pm$0.01	& 11.8     \\
NGC5746 &  14:44:55.98  & 01:57:18.1 & 0.28$\times$0.13 &  46.5 & 1.35$\pm$0.02 & 22.6	    \\
NGC5850 & 15:07:07.68  & 01:32:39.3  & 0.25$\times$0.13 &  45.4 & 0.10$\pm$0.01 & 11.1	   \\
NGC5921 & 15:21:56.49  & 05:04:14.3 & 0.22$\times$0.13 &  44.2 & 0.13$\pm$0.01	& 11.7  \\
NGC5985 &  15:39:37.06  & 59:19:55.2  & 0.21$\times$0.15 &  -86.4 &  0.53$\pm$0.01	 & 11.6    \\
NGC6340 &   17:10:24.84  & 72:18:15.9 & 0.19$\times$0.12 &  -41.8  &  0.59$\pm$0.01	& 10.2    \\
NGC6482 & 17:51:48.83  & 23:04:18.9 & 0.14$\times$0.14 &  34.5  & 0.42$\pm$	0.01 & 9.0   \\
NGC6703 & 18:47:18.82  & 45:33:02.3 & 0.14$\times$0.12 &  -20.7 & 0.63$\pm$0.01	& 9.2    \\
NGC6951 & 20:37:14.12  & 66:06:20.0 & 0.16$\times$0.12 &  24.0 & 0.26$\pm$0.01	& 11.2    \\
NGC7814 & 00:03:14.90  & 16:08:43.2  & 0.16$\times$0.14 &  -77.3 & 1.19$\pm$0.01 & 9.5	\\
IC356   & 04:07:46.89  & 69:48:44.7  & 0.25$\times$0.13 &  -81.8  & 1.28$\pm$0.01	& 16.2    \\
IC520   & 08:53:42.25  & 73:29:27.4  & 0.20$\times$0.13 &  -60.1 & 0.15$\pm$0.01	& 9.3   \\
\\
\hline
\hline
\\
\caption*{Note - Columns are: (1) galaxy name; (2) and (3) observed radio position (RA and Dec) of the detected source; (4) and (5) the beam size and beam position angle of the maps from which the previous columns data are derived; (6) peak 15 GHz radio flux density in mJy/beam with errors on the flux ; (7) rms in $\mu$Jy/beam}
\end{longtable}
\twocolumn

\pagebreak

\onecolumn
\begin{longtable}{llllllllllll}
\caption{Sample of LLAGN in the Palomar survey that were observed and reported in this paper (76 sources), as well as the previous radio detections obtained from the literature (68 sources).}\\ 
\hline
\hline
Name    & AType   & Dist & $\rm RA_{obs}$        & $\rm Dec_{obs}$      & $\rm S^{peak}_{15 GHz}$      & Det     & $\rm S^{total}_{15 GHz}$ & $\rm Log \, L^{peak}_{15 GHz}$   &  $\rm Log \, L^{total}_{15 GHz}$    & & \\
        &    & (Mpc)  & (h:m:s)        & (d:m:s)       & (mJy/beam)          &       & (mJy)     & (W/Hz)       &    (W/Hz)   &  \\
         \centering   (1)    & (2)  & (3)        & (4)       & (5)          & (6)      & (7)     & (8)       &    (9)   & (10) &  \\
        \hline
        \\
        \endhead    
NGC185  & S2      & 0.7  & --           & --          & --           & --      	& $<$1.1  & --   & $<$16.81     &  \\
NGC266  & L1.9    & 62.4 & 00:49:47.82 & 32:16:39.7 & 4.1          & N05  	& 4.1     & 21.28     &   21.28    &  \\
NGC315  & L1.9    & 65.8 & --           & --          & 470          & Lit 	&  --      & 23.38        &  --     &  \\
NGC410  & T2:     & 70.6 & 01:10:58.91  & 33:09:07.1 & 0.72 	     & S18   	& 0.7     & 20.64     &  20.62     &  \\
NGC428  & L2/T2:  & 14.9 & --           & --          & --           & --      	& $<$0.9  & --   & $<$19.38        &  \\
NGC474  & L2::    & 32.5 & --           & --          & --           & --      	& $<$1.5   & --  & $<$20.28       &  \\
NGC488  & T2::    & 29.3 & 01:21:46.80  & 05:15:24.6 & 0.14  	     & S18   	& 0.14    & 19.18     & 19.15      &  \\
NGC521  & T2/H:   & 67   & 01:24:33.77  & 01:43:52.9 & 0.34	     & S18   	& 0.32    & 20.27     &   20.23    &  \\
NGC660  & T2/H:   & 11.8 & 01:43:02.32  & 13:38:44.9 & 62.80	     & S18   	& 68.11   & 21.02     &   21.05    &  \\
NGC676  & S2:     & 19.5 & --           & --          & --           & --      	& $<$1.5  & --   & $<$19.83        &  \\
NGC718  & L2      & 21.4 & --           & --          & --           & --      	& $<$1.5     & -- & $<$19.91        &  \\
NGC777  & S2/L2:: & 66.5 & 02:00:14.91  & 31:25:45.9 & 0.54 	     & S18   	& 0.58    & 20.45      &   20.48    &  \\
NGC841  & L1.9:   & 59.5 & 02:11:17.37  & 37:29:49.7 & 0.06         & S18   	& 0.04 	  & 19.41  & 19.23 &  \\
NGC1055 & T2/L2:: & 12.6 & 02:41:45.20  & 00:26:38.1 & 0.18         & S18   	& 0.18 	  & 19.54 & 19.53 & \\
NGC1058 & S2      & 9.1  & --           & --          & --           & --      	& $<$0.9 & --    & $<$18.95      &  \\
NGC1167 & S2      & 65.3 & 03:01:42.35 & 35:12:20.1 & 44.9         & N05  	& 249     & 22.36     &  23.10     &  \\
NGC1169 & L2      & 33.7 & 03:03:34.74  & 46:23:11.1 & 0.36	     & S18   	& 0.37    & 19.70     &  19.70     &  \\
NGC1275 & S1.5    & 70.1 & --           & --          & 2970         & Lit 	& --      & 24.24        &   --    &  \\
NGC1961 & L2      & 53.1 & 05:42:04.65  & 69:22:42.4 & 0.50	     & S18   	& 0.53    & 20.23      &  20.25     &  \\
NGC2273 & S2      & 28.4 & 06:50:08.65 & 60:50:44.9 & 4.1          & N05  	& 5.1     & 20.60      &  20.69     &  \\
NGC2336 & L2/S2   & 33.9 & --           & --          & --           & --      	& $<$1.5 & --    & $<$20.31        &  \\
NGC2541 & T2/H:   & 10.6 & --           & --          & --           & --      	& $<$1   & --    & $<$19.13         &  \\
NGC2655 & S2      & 24.4 & 08:55:38.07 & 78:13:23.8 & 6            & N05  	& 6       & 20.63     &    20.63   &  \\
NGC2681 & L1.9    & 13.3 & 08:53:32.72  & 51:18:49.2 & 0.34	     & S18   	& 0.44    & 18.86     &  18.97     &  \\
NGC2683 & L2/S2   & 5.7  & 08:52:41.30  & 33:25:18.7 & 0.45	     & S18   	& 0.44    & 18.24     &  18.23     &  \\
NGC2768 & L2      & 23.7 & 09:11:37.41 & 60:02:14.9 & 7.9          & N05  	& 8.2     & 20.73     &   20.74    &  \\
NGC2787 & L1.9    & 13   & 09:19:18.61 & 69:12:11.6 & 7            & N05  	& 7       & 20.15     & 20.15      &  \\
NGC2832 & L2::    & 91.6 & 09:19:46.85  & 33:44:59.0 & 0.20	     & S18   	& 0.20    & 20.31     &  20.30     &  \\
NGC2841 & L2      & 12   & 09:22:02.68 & 50:58:35.7 & 1.1          & N05  	& 2.1     & 19.28     &   19.56    &  \\
NGC2859 & T2:     & 25.4 & 09:24:18.53  & 34:30:48.6 & 0.05         & S18   	& 0.04    & 18.61     &    18.49   &  \\
NGC2911 & L2      & 42.2 & 09:33:46.11 & 10:09:08.8 & 17.3         & N05 	& 17.7    & 21.57     &   21.57    &  \\
NGC2985 & T1.9    & 22.4 & 09:50:22.18  & 72:16:44.2 & 1.14	     & S18   	& 1.14    & 19.83     &    19.82   &  \\
NGC3031 & S1.5    & 3.6  & 09:55:33.17 & 69:03:55.1 & 164.1        & N05  	& 164.8   & 20.41     &   20.41    &  \\
NGC3079 & S2      & 20.4 & --           & --          & 48           & Lit 	& 59      & 21.38     &  21.47     &  \\
NGC3147 & S2      & 40.9 & 10:16:53.65 & 73:24:02.7 & 8            & N05  	& 8.1     & 21.20      &   21.21    &  \\
NGC3169 & L2      & 19.7 & 10:14:15.05 & 03:27:57.9 & 6.8          & N05  	& 6.8     & 20.50      &   20.50    &  \\
NGC3190 & L2      & 22.4 & 10:18:06.01 & 21:49:54.5 & 1.1          & N05  	& 0.8     & 19.82     &   19.68    &  \\
NGC3226 & L1.9    & 23.4 & 10:23:27.01 & 19:53:54.6 & 5            & N05  	& 5.4     & 20.51     &  20.55     &  \\
NGC3227 & S1.5    & 20.6 & 10:23:30.58 & 19:51:54.2 & 3.5          & N05  	& 4.7     & 20.25     &   20.38    &  \\
NGC3414 & L2      & 24.9 & 10:51:16.21 & 27:58:30.3 & 2.3          & N05  	& 2.4     & 20.23     &  20.25     &  \\
NGC3516 & S1.2    & 38.9 & --           & --          & 1.3          & Lit 	& 1.3     & 20.37     &   20.37    &  \\
NGC3607 & L2      & 19.9 & 11:16:54.67 & 18:03:06.4 & 1.6          & N05  	& 1.4     & 19.88     &  19.82     &  \\
NGC3627 & T2/S2   & 6.6  & 11:20:15.01 & 12:59:29.8 & 1.1          & N05  	& 2.9     & 18.76     &   19.18    &  \\
NGC3628 & T2      & 7.7  & 11:20:17.02 & 13:35:20.0 & 1.5          & N05  	& 2.2	& 19.02     & 19.19         &  \\
NGC3642 & L1.9    & 27.5 & 11:22:17.90   & 59:04:28.3 & 0.78	     & S18   	& 0.79    & 19.85     &    19.85   &  \\
NGC3675 & T2      & 12.8 & --           & --          & --           & --      	& $<$1       & --     &   $<$19.29    &  \\
NGC3718 & L1.9    & 17   & 11:32:34.86 & 53:04:04.5 & 10.5         & N05  	& 10.8    & 20.56     &   20.57    &  \\
NGC3735 & S2:     & 41   & 11:35:57.22  & 70:32:07.8 & 0.39	     & S18   	& 0.44    & 19.90     &  19.94     &  \\
NGC3780 & L2::    & 37.2 & 11:39:22.27 & 56:16:10.2 & 1.1          & N05  	& 1       & 20.26     &  20.22     &  \\
NGC3898 & T2      & 21.9 & 11:49:15.24  & 56:05:04.3 & 0.21	     & S18   	& 0.33    & 19.09        &   19.28    &  \\
NGC3900 & L2:     & 29.4 & --           & --          & --           & --      	& $<$1.5  & --   & $<$20.19      &  \\
NGC3917 & T2:     & 17   & --           & --          & --           & --      	& $<$1.2  & --   & $<$19.63        &  \\
NGC3941 & S2:     & 18.9 & --           & --          & --           & --      	& $<$1.1  & --   & $<$19.67   &  \\
NGC3945 & L2      & 22.5 & 11:53:13.61 & 60:40:32.1 & 1.8          & N05  	& 2.1     & 20.04     &    20.10   &  \\
NGC3953 & T2      & 17   & --           & --          & --           & --      	& $<$1.3  & --   & $<$19.64       &  \\
NGC3982 & S1.9    & 17   & 11:56:28.14  & 55:07:30.9 & 0.56	     & S18   	& 0.80       & 19.29     &  19.44     &  \\
NGC3992 & T2:     & 17   & 11:57:35.96  & 53:22:29.0 & 0.17	     & S18   	& 0.19    & 18.77     &  18.81     &  \\
NGC3998 & L1.9    & 21.6 & --           & --          & 57           & Lit 	& 57      & 21.50      &  21.50     &  \\
NGC4013 & T2      & 17   & --           & --          & --           & --      	& 1       & --     &   19.54    &  \\
NGC4036 & L1.9    & 24.6 & 12:01:26.75  & 61:53:44.6 & 0.60	     & S18   	& 0.85    & 19.64     &  19.79     &  \\
NGC4051 & S1.2    & 17   & 12:03:09.61  & 44:31:52.7 & 0.48         & S18   	& 0.48    & 19.22     &  19.22     &  \\
NGC4111 & L2      & 17   & 12:07:03.13  & 43:03:56.3 & 0.19	     & S18   	& 0.24    & 18.81     &  18.92     &  \\
NGC4125 & T2      & 24.2 & --           & --          & --           & --      	& $<$1  & --     & $<$19.85       &  \\
NGC4138 & S1.9    & 17   & 12:09:29.80 & 43:41:06.9 & 1.5          & N05  	& 1.3     & 19.71     &  19.65     &  \\
NGC4143 & L1.9    & 17   & 12:09:36.07 & 42:32:03.0 & 3.3          & N05  	& 3.3     & 20.06     & 20.06      &  \\
NGC4145 & T2:     & 20.7 & 12:10:01.56  & 39:53:00.9 & 0.09         & S18   	& 0.12     & 18.69     & 18.79      &  \\
NGC4150 & T2      & 9.7  & --           & --          & --           & --      	& $<$1    & --   & $<$19.05        &  \\
NGC4151 & S1.5    & 20.3 & --           & --          & 12.5         & Lit 	& 17.6    & 20.79     &  20.94     &  \\
NGC4168 & S1.9:   & 16.8 & 12:12:17.27 & 13:12:18.7 & 3            & N05  	& 3.1     & 20.00     &  20.02     &  \\
NGC4169 & S2      & 50.4 & 12:12:19.09 & 29:10:45.0 & 1.2          & N05  	& 1       & 20.56     &   20.48    &  \\
NGC4203 & L1.9    & 9.7  & 12:15:05.06 & 33:11:50.2 & 9.5          & N05  	& 9.5     & 20.03     &   20.03    &  \\
NGC4216 & T2      & 16.8 & 12:15:54.37 & 13:08:58.1 & 1.2          & N05  	& 1.3     & 19.61     &  19.64     &  \\
NGC4220 & T2      & 17   & --           & --          & --           & --      	& $<$1.4 & --    & $<$19.7        &  \\
NGC4258 & S1.9    & 6.8  & 12:18:57.51 & 47:18:14.3 & 2.6          & N05  	& 3       & 19.16     &   19.22    &  \\
NGC4261 & L2      & 35.1 & --           & --          & 300          & Lit 	& --      & 22.65     &    --   &  \\
NGC4278 & L1.9    & 9.7  & 12:20:06.82 & 29:16:50.7 & 88.3         & N05  	& 89.7    & 21.00        &  21.00     &  \\
NGC4293 & L2      & 17   & 12:21:12.81 & 18:22:56.7 & 0.7          & N05  	& 1.4     & 19.38     & 19.68      &  \\
NGC4314 & L2      & 9.7  & --           & --          & --           & --      	& $<$1    & --   & $<$19.05         &  \\
NGC4346 & L2::    & 17   & 12:23:27.96  & 46:59:37.6 & 0.06         & S18   	& 0.07    & 18.33     &  18.38     &  \\
NGC4374 & L2      & 16.8 & 12:25:03.74 & 12:53:13.2 & 180.7        & N05  	& 183.7   & 21.79     &   21.79    &  \\
NGC4388 & S1.9    & 16.8 & 12:25:46.78 & 12:39:43.8 & 2.2          & N05  	& 3.7     & 19.87     &   20.09    &  \\
NGC4395 & S1.8    & 3.6  & 12:25:48.88  & 33:32:48.7 & 0.17         & S18   	& 0.2     & 17.41     &   17.49    &  \\
NGC4414 & T2:     & 9.7  & --           & --          & --           & --      	& $<$0.9  & --   & $<$19.01   &  \\
NGC4419 & T2      & 16.8 & 12:26:56.45 & 15:02:50.8 & 2.7          & N05  	& 3.6     & 19.96     &   20.08    &  \\
NGC4450 & L1.9    & 16.8 & 12:28:29.59 & 17:05:06.0 & 2            & N05  	& 2.7     & 19.83     &  19.96     &  \\
NGC4472 & S2::    & 16.8 & 12:29:46.76 & 08:00:01.7 & 3.7          & N05  	& 4.1     & 20.10      &  20.14     &  \\
NGC4486 & L2      & 16.8 & 12:30:49.42 & 12:23:28.0 & 2725.7       & N05  	& 2835.7  & 22.96     &  22.98     &  \\
NGC4548 & L2      & 16.8 & 12:35:26.45 & 14:29:46.7 & 1.2          & N05  	& 1.2     & 19.61     &   19.61    &  \\
NGC4550 & L2      & 16.8 & 12:35:30.08 & 12:13:18.6 & 0.7          & N05  	& 0.7     & 19.37     &   19.37    &  \\
NGC4552 & T2:     & 16.8 & 12:35:39.81 & 12:33:22.8 & 58.1         & N05  	& 58.6    & 21.29     &  21.29     &  \\
NGC4565 & S1.9    & 9.7  & 12:36:20.78 & 25:59:15.6 & 3.7          & N05  	& 3.7     & 19.62     &    19.62   &  \\
NGC4579 & S1.9/L  & 16.8 & 12:37:43.52 & 11:49:05.5 & 27.6         & N05  	& 28.3    & 20.97     &  20.98     &  \\
NGC4589 & L2      & 30   & 12:37:24.99 & 74:11:30.9 & 11.7         & N05  	& 11.9    & 21.10      &  21.11     &  \\
NGC4636 & L1.9    & 17   & 12:42:49.83 & 02:41:15.9 & 1.6          & N05  	& 1.9     & 19.74     &  19.82     &  \\
NGC4736 & L2      & 4.3  & 12:50:53.07 & 41:07:12.9 & 1.9          & N05  	& 1.7     & 18.62     &  18.57     &  \\
NGC4750 & L1.9    & 26.1 & 12:50:07.32  & 72:52:28.6 & 0.86	     & S18   	& 0.95    & 19.84     &  19.89     &  \\
NGC4762 & L2:     & 16.8 & 12:52:55.90 & 11:13:46.6 & 0.9          & N05  	& 1.3     & 19.48     &   19.64    &  \\
NGC4772 & L1.9    & 16.3 & 12:53:29.16 & 02:10:06.2 & 3.3          & N05  	& 3.4     & 20.02     &    20.03   &  \\
NGC5033 & S1.5    & 18.7 & 13:13:27.47 & 36:35:37.9 & 1.4          & N05  	& 1.4     & 19.77     & 19.77      &  \\
NGC5194 & S2      & 7.7  & 13:29:52.71  & 47:11:42.8 & 0.25	     & S18   	& 0.38    & 18.25     &   18.43    &  \\
NGC5195 & L2:     & 9.3  & 13:29:59.53  & 47:15:58.3 & 0.25	     & S18   	& 0.37    & 18.41     & 18.58      &  \\
NGC5273 & S1.5    & 21.3 & 13:42:08.38  & 35:39:15.5 & 0.19         & S18   	& 0.37    & 19.01     &  19.30     &  \\
NGC5297 & L2      & 37.8 & --           & --          & --           & --      	& $<$1.5   & --  & $<$20.41      &  \\
NGC5322 & L2::    & 31.6 & 13:49:15.26 & 60:11:25.9 & 12.7         & N05  	& 12.7    & 21.18     &  21.18     &  \\
NGC5353 & L2/T2:  & 37.8 & 13:53:26.69 & 40:16:58.9 & 18.3         & N05  	& 18.7    & 21.49      &  21.50     &  \\
NGC5354 & T2/L2:  & 32.8 & --           & --          & 9.7          & Lit 	& 9.7     & 21.09      &   21.09    &  \\
NGC5363 & L2      & 22.4 & 13:56:07.21 & 05:15:17.2 & 38.1         & N05 	& 40.7    & 21.36     &    21.39   &  \\
NGC5371 & L2      & 37.8 & --           & --          & --           & --      	& $<$1.5 & --    & $<$20.41     &  \\
NGC5377 & L2      & 31   & 13:56:16.83 & 47:14:06.3 & 3            & N05  	& 3.1     & 20.54     &   20.55    &  \\
NGC5395 & S2/L2   & 46.7 & 13:58:37.96  & 37:25:28.2 & 0.11	     & S18   	& 0.11    & 19.48     &   19.46    &  \\
NGC5448 & L2      & 32.6 & 14:02:50.05  & 49:10:21.2 & 0.06         & S18   	& 0.16    & 18.87     &  19.31     &  \\
NGC5485 & L2:     & 32.8 & 14:07:11.35  & 55:00:06.0 & 0.56	     & S18   	& 0.62    & 19.86     &  19.90     &  \\
NGC5566 & L2      & 26.4 & 14:20:19.89  & 03:56:01.4 & 1.40  	     & S18   	& 1.66    & 20.07     &   20.14    &  \\
NGC5631 & S2/L2:  & 32.7 & 14:26:33.29  & 56:34:57.4 & 0.64	     & S18   	& 0.64    & 19.91     &   19.91    &  \\
NGC5656 & T2::    & 42.6 & --           & --          & --           & --      	& $<$1  & --     & $<$20.34      &  \\
NGC5678 & T2      & 35.6 & --           & --          & --           & --      	& $<$1   & --    & $<$20.18         &  \\
NGC5701 & T2:     & 26.1 & --           & --          & --           & --      	& $<$1.1   & --  & $<$19.95     &  \\
NGC5746 & T2      & 29.4 & 14:44:55.98  & 01:57:18.1 & 1.35	     & S18   	& 1.40    & 20.14     &   21.62    &  \\
NGC5813 & L2:     & 28.5 & 15:01:11.23 & 01:42:07.1 & 2.2          & N05  	& 2.4     & 20.33     &  20.37     &  \\
NGC5838 & T2::    & 28.5 & --           & --          & 1.6          & Lit 	& 1.6     & 20.19     &   20.19    &  \\
NGC5846 & T2:     & 28.5 & --           & --          & 6.3          & Lit 	& 6.3     & 20.79     &  20.79     &  \\
NGC5850 & L2      & 28.5 & 15:07:07.68  & 01:32:39.3 & 0.10	     & S18   	& 0.10    & 19.00        & 18.99      &  \\
NGC5866 & T2      & 15.3 & 15:06:29.50 & 55:45:47.6 & 7.1          & N05  	& 7.5     & 20.30      &   20.32    &  \\
NGC5879 & T2/L2   & 16.8 & --           & --          & --           & --      	& $<$1.1  & --   & $<$19.57         &  \\
NGC5921 & T2      & 25.2 & 15:21:56.49  & 05:04:14.3 & 0.13	     & S18   	& 0.15    & 19.01     &  19.05     &  \\
NGC5970 & L2/T2:  & 31.6 & --           & --          & --           & --      	& --      & --        &   --    &  \\
NGC5982 & L2::    & 38.7 & --           & --          & --           & --     	& --      & --        &   --    &  \\
NGC5985 & L2      & 39.2 & 15:39:37.06  & 59:19:55.2 & 0.53	     & S18   	& 0.54    & 19.99        &   19.99    &  \\
NGC6340 & L2      & 22   & 17:10:24.84  & 72:18:15.9 & 0.59	     & S18   	& 0.61    & 19.53     &    19.55   &  \\
NGC6384 & T2      & 26.6 & --           & --          & --           & --      	& $<$1   & --    & $<$19.93         &  \\
NGC6482 & T2/S2:: & 52.3 & 17:51:48.83  & 23:04:18.9 & 0.42	     & S18   	& 0.45    & 20.14     &  20.18     &  \\
NGC6500 & L2      & 39.7 & 17:55:59.782 & 18:20:17.6 & 83.5         & N05  	& 85      & 22.19      &  22.20     &  \\
NGC6503 & T2/S2:  & 6.1  & --           & --          & --           & --     	& $<$1   & -- & $<$18.65            &  \\
NGC6703 & L2::    & 35.9 & 18:47:18.82  & 45:33:02.3 & 0.63	     & S18   	& 0.63    & 19.99     &  19.99     &  \\
NGC6951 & S2      & 24.1 & 20:37:14.12  & 66:06:20.0 & 0.26	     & S18   	& 0.28    & 19.26     & 19.29      &  \\
NGC7479 & S1.9    & 32.4 & 23:04:56.63 & 12:19:22.6 & 2.4          & N05  	& 2.5     & 20.48     & 20.49      &  \\
NGC7626 & L2::    & 45.6 & 23:20:42.54 & 08:13:00.9 & 39.8         & N05  	& 40      & 21.99     &	21.99       &  \\
NGC7743 & S2      & 24.4 & 23:44:21.36 & 09:56:03.8 & 1            & Lit  	& 1.5     & 19.85     &  20.03     &  \\
NGC7814 & L2::    & 15.1 & 00:03:14.90  & 16:08:43.2 & 1.19	     & S18   	& 1.19    & 19.51     &  19.51     &  \\
IC239   & L2::    & 16.8 & --           & --          & --           & --      	& --      & --        &   --    & \\
IC356   & T2      & 18.1 & 04:07:46.89  & 69:48:44.7 & 1.28	     & S18   	& 1.23    & 19.70        &   19.68    &  \\
IC520   & T2:     & 47.0 & 08:53:42.25  & 73:29:27.4 & 0.15	     & S18   	& 0.15    & 19.60        & 19.59      &  \\
IC1272  & T2/L2   & 8.2  & --           & --          & --           & --      	& --      & --        &   --    &  \\
\caption*{Columns are: (1) galaxy name; (2) nuclear activity type from \citet{h97} (`L' : LINER, `S' : Seyfert, `T' : transition objects; `2' : no broad H$\alpha$, `1.9' : broad H$\alpha$ is present, but no broad H$\beta$, `1.0' or `1.5' : both broad H$\alpha$ and broad H$\beta$, `:' : uncertain classifications, `::' : highly uncertain classifications); (3) distance to the galaxy in Mpc; (4) and (5) are the 2 cm radio position; (6) peak flux of the nuclear radio source; (7) the source from where the radio values are taken (S18 : observed and detected in this survey, -- : observed in this survey, but not detected, N05 : reported in Nagar et al. 2005, Lit : compiled by Nagar et al. 2005 from literature); (8) integrated flux density of the nuclear source; (9) and (10) are the radio luminosity of the source (derived from the peak and total radio flux-density, respectively).}
\end{longtable}
\twocolumn

\nopagebreak

\onecolumn
\begin{longtable}{llllll}
\caption{Summary of the Palomar LLAGN sample with nuclear radio detections at 15 GHz in high resolution (mainly from this paper, and \cite{n05}), with a compilation of mass and X-ray luminosity from the literature (references listed at the end of the table).}\\ 
\hline
\hline
Name    & Log L$_R$   & Log L$_X$ & Log M   & Dist  & Ref\\
        &        (erg/s)  & (erg/s)        & (M$_{\odot}$)       & (Mpc)       &      \\
        \\
            (1)    & (2)  & (3)        & (4)       & (5)       & (6)   \\
        \hline
        \\
        \endhead 
NGC266   & 38.46 & 40.77 & 8.60  & 62.4 & 5\\
NGC315   & 40.56 & 41.77 & 9.22 & 65.8 & 2\\
NGC410   & 37.81 & 40.71 & 9.19 & 70.6  & 2\\
NGC660   & 38.20  & 38.32 & 7.30  & 11.8 & 4\\
NGC1055  & 35.72 & 38.38 & 6.25 & 12.6 & 4\\
NGC1167  & 39.54 & 39.60  & 8.47 & 65.3 & 1\\
NGC1275  & 41.42 & 43.40  & 8.86 & 70.1 & 6\\
NGC2273  & 37.77 & 40.93 & 7.63 & 28.4 & 1\\
NGC2655  & 37.81 & 40.89 & 7.79 & 24.4 & 1\\
NGC2681  & 36.03 & 39.27 & 6.94 & 13.3 & 2 \\
NGC2683  & 35.42 & 38.21 & 7.34 & 5.7  & 7\\
NGC2787  & 37.33 & 38.80  & 8.31 & 13.0 & 2\\
NGC2841  & 36.45 & 39.22 & 8.52 & 12.0 & 2 \\
NGC3031  & 37.58 & 39.48 & 7.82 & 3.6  & 1\\
NGC3079  & 38.55 & 40.26 & 8.08 & 20.4 & 1\\
NGC3147  & 38.38 & 41.49 & 8.50  & 40.9 & 1\\
NGC3169  & 37.68 & 41.35 & 8.16 & 19.7 & 8\\
NGC3226  & 37.69 & 40.80  & 8.22 & 23.4 & 2\\
NGC3227  & 37.43 & 41.64 & 7.43 & 20.6 & 1\\
NGC3414  & 37.41 & 39.86 & 8.67 & 24.9 & 2\\
NGC3516  & 37.55 & 42.42 & 8.07 & 38.9 & 1\\
NGC3607  & 37.06 & 38.76 & 8.62 & 19.9 & 2\\
NGC3627  & 35.93 & 39.41 & 7.23 & 6.6  & 2\\
NGC3628  & 36.20 & 39.94 & 6.18 & 7.7  & 2\\
NGC3718  & 37.74 & 41.06 & 7.77 & 17.0 & 5\\
NGC3735  & 37.08 & 40.53 & 7.51 & 41.0 & 1\\
NGC3945  & 37.21 & 39.12 & 8.19 & 22.5 & 2\\
NGC3982  & 36.47 & 39.78 & 6.05 & 17.0 & 1\\
NGC3998  & 38.68 & 41.32 & 9.23 & 21.6 & 2\\
NGC4036 & 36.82 & 38.12 & 8.45 & 24.6 & 2\\
NGC4051  & 36.40  & 41.35 & 6.49 & 17.0 & 1\\
NGC4111  & 35.99 & 40.36 & 7.62 & 17.0 & 2\\
NGC4138  & 36.89 & 41.28 & 7.17 & 17.0 & 1\\
NGC4143  & 37.23 & 40.03 & 8.34 & 17.0 & 8\\
NGC4151  & 37.97 & 43.01 & 6.68 & 20.3 & 1\\
NGC4168  & 37.18 & 39.30  & 8.10  & 16.8 & 1\\
NGC4169  & 37.74 & 40.84 & 8.09 & 50.4 & 1\\
NGC4203  & 37.21 & 40.37 & 7.89 & 9.7  & 3\\
NGC4258  & 36.33 & 40.90  & 7.62 & 6.8  & 3\\
NGC4261  & 39.82 & 41.07 & 9.26 & 35.1 & 2\\
NGC4278  & 38.17 & 39.23 & 8.88 & 9.7  & 2\\
NGC4374  & 38.96 & 39.53 & 9.25 & 16.8 & 2\\
NGC4388  & 37.05 & 41.85 & 6.56 & 16.8 & 1\\
NGC4395  & 34.59 & 40.08 & 4.07 & 3.6  & 1\\
NGC4450  & 37.01 & 40.35 & 7.42 & 16.8 & 3\\
NGC4472  & 37.27 & 39.85 & 9.12 & 16.8 & 1\\
NGC4486  & 40.14 & 40.82 & 9.42 & 16.8 & 2\\
NGC4548  & 36.78 & 39.79 & 7.03 & 16.8 & 8\\
NGC4552  & 38.47 & 39.25 & 8.81 & 16.8 & 2\\
NGC4565  & 36.80 & 41.32 & 7.43 & 9.7  & 1\\
NGC4579  & 38.15 & 41.17 & 7.86 & 16.8 & 2\\
NGC4589  & 38.28 & 38.91 & 8.54 & 30.0 & 2\\
NGC4636  & 36.92 & 39.03 & 8.32 & 17.0 & 2\\
NGC4736  & 35.80  & 38.59 & 7.00  & 4.3  & 2\\
NGC4750  & 37.02 & 40.08 & 7.43 & 26.1 & 5\\
NGC4772  & 37.20  & 39.85 & 7.62 & 16.3 & 5\\
NGC5033  & 36.94 & 42.64 & 7.67 & 18.7 & 1\\
NGC5194  & 35.43 & 39.30  & 6.66 & 7.7  & 1\\
NGC5273  & 36.18 & 41.58 & 5.99 & 21.3 & 1\\
NGC5363  & 38.54 & 39.78 & 8.57 & 22.4 & 2\\
NGC5746  & 37.32 & 40.22 & 8.29 & 29.4 & 2\\
NGC5813  & 37.51 & 38.80  & 8.68 & 28.5 & 2\\
NGC5838  & 37.37 & 39.20  & 8.92 & 28.5 & 2\\
NGC5846  & 37.96 & 40.81 & 8.67 & 28.5 & 2\\
NGC5866  & 37.47 & 38.29 & 7.92 & 15.3 & 2\\
NGC6482  & 37.32 & 39.33 & 9.27 & 52.3 & 2\\
NGC6500  & 39.37 & 40.11 & 8.44 & 39.7 & 8\\
NGC6951  & 36.44 & 39.48 & 7.29 & 24.1 & 1\\
NGC7479  & 37.66 & 40.45 & 7.72 & 32.4 & 1\\
NGC7743  & 37.03 & 37.85 & 6.50  & 24.4 & 1\\
\\
\hline
\hline
\\
\caption*{Note - Columns are: (1) galaxy name; (2) logarithmic value of radio luminosity at 15 GHz (in erg/s) with the VLA; (3) logarithmic value of X-ray luminosity at 2-10 keV (in erg/s), taken from Literature (reference given in column 6); (4) logarithmic value of black hole mass (in M$_{\odot}$), estimated from stellar velocity dispersions reported in \citet{h09} using the empirical relation given in \citet{m11}; (5) distance to the source in Mpc; (6) literature references for the X-ray luminosity at 2-10 KeV range quoted in column (3)\\

References for X-ray luminosity : 1: \cite{aky2009}, 2 : \cite{gon2009}, 3 : \cite{ter2002}, 4 : \cite{filho2004}, 5 : \cite{yo}, 6 : \cite{a2001}, 7 : \cite{pan2006}, 8 : \cite{ter2003}}
\end{longtable}
\twocolumn

\nopagebreak 
 
%%%


\begin{thebibliography}{dummy}


\bibitem[Akylas \& Georgantopoulos(2009)]{aky2009} 
Akylas, A. \& Georgantopoulos, I. 2009, A\&A, 500, 999

\bibitem[Allen et al.(2001)]{a2001} 
Allen S. W., Fabian A. C., Johnstone R. M., Arnaud K. A. \& Nulsen P. E. J., 2001, MNRAS, 322, 589

\bibitem[Argo et al.(2015)]{argo2015} 
Argo M. K., van Bemmel I. M., Connoll S. D. \& Beswick R. J., 2015, MNRAS, 452, 1081

\bibitem[Baldi et al.(2018)]{lemm2018}
Baldi R. D., Williams D. R. A, McHardy I. M., Beswick R. J., Argo M. K., et al. 2018, MNRAS, 476, 3, 3478

\bibitem[Balmaverde et al.(2016)]{extra2006}
Balmaverde B., Capetti A. \& Grandi P., 2006, A\&A, 451, 35

\bibitem[Blandford \& K{\" o}nigl(1979)]{bk79}
Blandford R. D. \& K{\" o}nigl A., 1979, ApJ, 232, 34

\bibitem[Carral et al.(1990)]{caret90}
Carral, P., Turner, J.~L., \& Ho, P.~T.~P.\ 1990, \apj, 362, 434 

\bibitem[Corbel et al.(2008)]{c08}
Corbel S., K{\" o}rding E., Kaaret P., 2008, MNRAS, 389, 1697

\bibitem[Corbel et al.(2013)]{c13}
Corbel S., Coriat M., Brocksopp C., Tzioumis A. K., Fender R. P., Tomsick J. A., Buxton M. M., Bailyn C. D., 2013, MNRAS, 428, 2500

\bibitem[Fabbiano et al.(1989)]{fabet89}
Fabbiano, G., Gioia, I. M., \& Trinchieri, G. 1989, \apj, 374, 127

\bibitem[Falcke \& Biermann(1995)]{fb99}
Falcke H. \& Biermann P. L., 1999, A\&A, 342, 49

\bibitem[Falcke et al.(2000)]{heino2000}
Falcke H., Nagar N. M., Wilson A. S. \& Ulvestad J. S., 2000, \apj, 542, 197

\bibitem[Falcke (2001)]{f2001}
Falcke H., 2001, in Schielicke R. E., ed., Reviews in Modern Astronomy 14: Dynamic Stability and Instabilities in the Universe . Astronomische Gesellschaft, Hamburg, Germany, p. 15

\bibitem[Falcke, K{\" o}rding, \& Markoff(2004)]{f04}
Falcke, H., K{\" o}rding, E., \& Markoff, S.\ 2004, \aap, 414, 895 

\bibitem[Fender, Belloni \&Gallo(2004)]{fbg04}
Fender R. P., Belloni T. M. \& Gallo E., 2004, MNRAS, 355, 1105

\bibitem[Fanaroff \& Riley(1974)]{fr}
Fanaroff B.L. \& Riley J.M., 1974, MNRAS, 167, 31P

\bibitem[Filho et al.(2002)]{filho2002}
Filho, M. E., Barthel, P. D., \& Ho, L. C. 2002, \apjs, 142, 223

\bibitem[Filho et al.(2004)]{filho2004} 
Filho, M.~E., Fraternali, F., Markoff, S., Nagar, N.~M., Barthel, P.~D., Ho, L.~C., \& Yuan, F., 2004, \aap, 418, 429 

\bibitem[Filho et al.(2006)]{filho2006} 
Filho, M.~E., Barthel, P.~D. \& Ho, L.~C., 2006, \aap, 451, 71

\bibitem[Gallo et al.(2006)]{g06} 
Gallo E., Fender R. P., Miller-Jones J. C. A., Merloni A., Jonker P. G., Heinz S., Maccarone T. J., van der Klis M., 2006, MNRAS, 370, 13511360

\bibitem[Gonz\'alez-Mart\'in et al.(2009)]{gon2009} 
Gonz\'alez-Mart\'in, O., Masegosa, J., Márquez, I., Guainazzi, M., \& Jim\'enez-Bailon, E. 2009, A\&A, 506, 1107

\bibitem[Heckman et al.(2005)]{h05} 
Heckman T. M., Ptak A., Hornschemeier A., Kauffmann G., 2005, ApJ, 634,161

\bibitem[Ho et al.(1995)]{h95}
Ho L. C., Filippenko A. V., \& Sargent W. L. W. 1995, \apjs, 98, 477

\bibitem[Ho et al.(1997)]{h97}
Ho L. C., Filippenko A. V., \& Sargent W. L. W. 1997, \apjs, 112, 315 

\bibitem[Ho (1999)]{ho99}
Ho L. C., 1999, ApJ, 516, 672

\bibitem[Ho, Filippenko, \& Sargent(2003)]{hoet03} 
Ho L. C., Filippenko A. V., \& Sargent W. L. W.  2003, \apj, 583, 159 

\bibitem[Ho(2008)]{ho2008}
Ho L. C., 2008, ARA\&A, 46, 475

\bibitem[Ho et al.(2009)]{h09}
Ho L. C., Greene J. E., Filippenko A. V., Sargent W. L. W., 2009, ApJS, 183, 1

\bibitem[Ho et al.(2010)]{hoet2010}
Ho L. C., Rudnick G., Rix H.-W., Shields J. C., McIntosh D. H., Filippenko A. V., Sargent W. L. W. \& Eracleous M., 2000, ApJ, 541, 120

\bibitem[Ho \& Ulvestad(2001)]{houlv01} 
Ho L. C. \& Ulvestad, J. S., 2001, \apjs, 133, 77 

\bibitem[Hummel et al.(1987)]{humet87}
 Hummel E., van der Hulst J. M., Keel W.~C., \& Kennicutt R.~C.\ 1987, \aaps, 70, 517 
 
 \bibitem[K{\" o}rding et al.(2006)]{elmar06}
K{\" o}rding E. G., Fender R. P. \& Migliari S., 2006, MNRAS , 369, 1451
 
%\bibitem[Kukula et al.(1995)]{kukula1995}
%Kukula M. J., Pedlar A., Baum S. A. \& O'Dea C. P., 1995, MNRAS, 276, 1262

 %\bibitem[Laurent-Muehleisen et al.(1997)]{lauet97}
%Laurent-Muehleisen, S. A., Kollgaard, R. I., Ryan, P. J., Feigelson, E. D., Brinkmann, W., \& Siebert, J., 1997, \aaps, 122, 235

\bibitem[\protect\citeauthoryear{Li, Wu \& Wang}{2008}]{lww08}
Li Z, Wu XB \& Wang R., 2008, ApJ, 688, 826

\bibitem[Markoff et al.(2001)]{sara2001} 
Markoff S., Falcke H. \& Fender R., 2001, A\&A, 372, L25

\bibitem[McConnell et al.(2011)]{m11} 
McConnell N. J., Ma C.-P., Gebhardt K., Wright S. A., Murphy J. D., Lauer T. R., Graham J. R., Richstone D. O., 2011, Nature, 480, 215

\bibitem[McMullin et al.(2007)]{casa} 
McMullin J. P., Waters B., Schiebel D., Young W. \& Golap K., 2007, Astronomical Data Analysis Software and Systems XVI (ASP Conf. Ser. 376), ed. R. A. Shaw, F. Hill \& D. J. Bell (San Francisco, CA: ASP), 127

\bibitem[Merloni, Heinz, \& di Matteo(2003)]{m03} 
Merloni, A., Heinz, S., \& di Matteo, T.\ 2003, \mnras, 345, 1057 

%\bibitem[Meurs \& Wilson(1984)]{mw94}
%Meurs E. J. A. \& Wilson A. S., 1984, A\&A, 136, 206 

\bibitem[Mohan \& Rafferty(2015)]{pybdsm}
Mohan N. \& Rafferty D., 2015, Astrophysics Source Code Library, 1502.007

\bibitem[Nagar et al.(2005)]{n05}
Nagar, N. M., Falcke \& H., Wilson., 2005, \aa, 435 

\bibitem[Panessa et al.(2006)]{pan2006}
Panessa, F., Bassani, L., Cappi, M., Dadina, M., Barcons, X., Carrera, F. J., Ho, L. C., \& Iwasawa, K., 2006, A\&A, 455, 173

\bibitem[Panessa \& Giroletti(2013)]{panessa2013}
Panessa F. \& Giroletti M., 2013, MNRAS, 432, 2

\bibitem[Perley et al.(2011)]{vla11}
Perley R. A., Chandler C. J., Butler B. J. \& Wrobel J. M., 2011, ApJ, 739, L1

\bibitem[Quataert et al.(1999)]{q99}
Quataert E., Narayan R. \& Reid M. J., 1999, ApJ, 517, 2

\bibitem[Saikia, K{\" o}rding \& Falcke(2015)]{s15}
Saikia P., K{\" o}rding E. \& Falcke H., 2015, MNRAS, 450, 2317

\bibitem[Saikia, K{\" o}rding \& Dibi(2018)]{saikia18}
Saikia P., K{\" o}rding E. \& Dibi S., 2018, MNRAS, 477, 2119

\bibitem[Sikora et al.(2007)]{sikora2007}
Sikora M., Stawarz L. \& Lasota J. P., 2007, ApJ, 658, 815

\bibitem[Terashima et al.(2002)]{ter2002}
Terashima, Y., Iyomoto, N., Ho, L. C., \& Ptak, A. F., 2002, ApJS, 139, 1

\bibitem[Terashima \& Wilson(2003)]{ter2003}
Terashima, Y. \& Wilson, A. S., 2003, ApJ, 583, 145

%\bibitem[Thompson et al.(1980)]{thoet80} Thompson, A. R., Clark, B. G.,
%Wade C. M., \& Napier, P. J. 1980, \apjs, 44, 151     

\bibitem[Ulvestad \& Ho(2001)]{ulvho2001} 
Ulvestad, J.~S.~\& Ho, L.~C.\ 2001, \apjl, 562, L133

\bibitem[Wang et al.(2006)]{wang06}
Wang R., Wu XB \& Kong M., 2006, ApJ, 645, 890

\bibitem[White et al.(1997)]{w97}
White, R. L., Becker, R. H., Helfand, D. J., \& Gregg, M. D., 1997, \apj, 475, 479

\bibitem[Younes et al.(2011)]{yo}
Younes, G., Porquet, D., Sabra, B., \& Reeves, J. N., 2011, A\&A, 530, A149
 
\end{thebibliography}
\end{document}